\documentclass[3p,11pt,sort&compress]{elsarticle}

\usepackage{natbib}
\usepackage{algorithm, algorithmic}
\usepackage{enumerate}
\usepackage{amsmath, amsthm, amssymb, bm}
\usepackage{appendix}
\usepackage{graphicx}
\usepackage{wrapfig}
\usepackage{multirow}
\usepackage{caption}
\usepackage{subcaption}

\usepackage{afterpage}
\usepackage{hyperref}
\usepackage{url}
\usepackage{float}
\usepackage{textcomp}

\theoremstyle{definition}
\newtheorem{definition}{Definition}[section]

\makeatletter
\newcommand\footnoteref[1]{\protected@xdef\@thefnmark{\ref{#1}}\@footnotemark}
\makeatother

\begin{document}

\begin{frontmatter}

\title{Uncovering Urban Mobility and City Dynamics from Large-Scale Taxi Origin-Destination (O-D) Trips: Case Study in Washington DC Area}

\journal{{\em \textcopyright WIOMAX} Technical Report No. WIO-TR-18-003}

\author{Xiao-Feng Xie}
\ead{xie@wiomax.com}
\author{Zunjing Jenipher Wang}
\ead{wang@wiomax.com}
 
\address{WIOMAX LLC, PO Box 540, Rockville, MD 20848}

\vspace{10 mm}

\begin{abstract}
\label{pap:abstract}
We perform a systematic analysis on the large-scale taxi trip data to uncover urban mobility and city dynamics in multimodal urban transportation environments. As a case study, we use the taxi origin-destination trip data and some additional data sources in Washington DC area. We first study basic characteristics of taxi trips, then focus on five important aspects. Three of them concern urban mobility, which are respectively mobility and cost including effect of traffic congestion, trip safety, and multimodal connectivity; the other two pertain to city dynamics, which are respectively transportation resilience and the relation between trip patterns and land use. For these aspects, we use appropriate statistical methods and geographic techniques to mine patterns and characteristics from taxi trip data for better understanding qualitative and quantitative impacts of the inputs from key stakeholders on available measures of effectiveness on urban mobility and city dynamics, where key stakeholders include road users, system operators, and city. Finally, we briefly summarize our findings and discuss some critical roles and implications of the uncovered patterns and characteristics from the relation between taxi system and key stakeholders. The results can support road users by providing evidence-based information of trip cost, mobility, safety, multimodal connectivity and transportation resilience, can assist taxi drivers and operators to deliver transportation services in a higher quality of mobility, safety and operational efficiency, and can also help city planners and policy makers to transform multimodal transportation and to manage urban resources in a more effective and better way.
\end{abstract}


\end{frontmatter}

\vspace{1 mm}

\section{Introduction}

More than half of the world's population lives in cities now, and estimated 65-70\% of the world's population
will live in cities by 2050 \citep{UNurbanization2014}. The swell in urban populations puts significant pressure on urban transportation systems, and there has been increasing interest in gaining a better understanding on the urban mobility and city dynamics for achieving sustainable urban transportation systems \citep{litman2006issues,arribas2014accidental}. Accompanied by the city development and transformation, the taxi industry, as one of essential mode in multimodal urban transportation systems, has kept growing with a fast-paced evolution in many recent years. 

Serving diverse transportation demands from a large number of people and often equipped with Global Positioning System (GPS) devices, taxi cabs become participatory sensors generating a huge amount of data with massive spatio-temporal information on human activity and mobility. Extensive research has been conducted using taxi trip data ~\citep{castro2013taxi,ferreira2013visual}. Some work focused on understanding fundamental human mobility patterns and statistics. \cite{wang2015comparative} studied the common regularity of intra-city human mobility by taxi through a empirical comparative analysis. \citet{peng2012collective} found that people's travels on workdays using taxi follow a few primary flow patterns. \cite{tang2015uncovering} use observed origin-destination (OD) matrix to model the traffic distribution patterns. 

Other studies drew attentions to urban mobility and city dynamics. On urban mobility, most research aimed to extract traffic and transportation dynamics through data analysis of taxi trips, as according to which beneficial applications could be found and provided to road users or stakeholders. \citet{geroliminis2008existence} revealed urban-scale macroscopic fundamental diagrams using taxi trips. \citet{zhan2013urban} presented a method to estimate possible linked travel time in urban road networks using the OD trip data of taxis. \citet{yang2017scalable} proposed a smart driving direction and route planning system to leverage the experience of taxi drivers on choosing driving directions. \cite{schaller2005regression} and \cite{zhang2016impacts} found that taxi services are well mutual complementary with other transportation modes such as airport and public transits in multimodal urban traffic environments. \citet{donovan2017empirically} proposed quantitative methods measuring the resilience of transportation systems, by using the deviations of historical distribution in normalized travel time of taxi among various regions of a city. Regarding traffic congestion --- a significant problem in major cities \citep{schrank20152015}, a few studies \citep{gan2013optimal,qian2017time,yuan2017modeling} presented pricing schemes to support offering taxi drivers extra incentives so that they would work in congested peak hours. Concerning road safety, \citet{boufous2009factors} and \citet{wu2016discrepancy} investigated some safety issues of taxi trips.

On city dynamics, researchers targeted to gain more insights on social dynamics and urban structures \citep{castro2013taxi,anas1998urban}. Ref.~\cite{qian2015spatial} revealed that urban forms have significant impact on urban taxi ridership. \citet{liu2012understanding} interpreted the spatial and temporal patterns observed in intra-urban human mobility through integrating spatial heterogeneity and distance decay of trips. \citet{shen2017discovering} and \citet{zhang2017revealing} used taxi data to identify points of interest (POIs) and statistically significant spatial clusters based on hot spot analysis. \citet{guo2012discovering} detected location characteristics and spatial structures from spatial and temporal patterns in the taxi trip movements. \citet{liu2012urban} linked the temporal variations in taxi pick-ups and drop-offs with various land use features. \citet{liu2015revealing} identified sub-regional structures by exploring the inherent connection between travel patterns and city structure. \citet{zhang2017revealing} revealed intra-urban travel patterns and service ranges by analyzing taxi trajectories. \citet{zhu2017street} clustered street types based on their dynamic functions. 

It is important to understand multimodal urban mobility from the viewpoint of taxi services. The significance is threefold. First, taxicabs are mixed with other motor vehicles in traffic flows and face the same traffic congestion problem on urban road networks. Thus the taxi trip data can serve as suitable source for measuring the impact of urban congestion on mobility and trip cost. Notice that the mobility is similar between taxicabs and passenger vehicles as they are mixed in traffic, whereas the trip cost of using taxi services is normally higher than that of using public transit services. Second, taxi services, along with the broader ridesharing services \citep{cramer2016disruptive}, represent a class of shared-use mobility with the for-hire vehicles serving the travelers who do not own or use their private vehicles. As traffic congestion and vehicle emissions are two serious problems for most major cities nowadays, there is an increasing need to encourage the less use of private vehicles~\citep{beirao2007understanding,jittrapirom2017mobility} for reducing the total vehicle miles of travel (VMT) and making urban transportation to be more sustainable. Understanding the role of taxi services in multimodal transportation is essential for fostering the process of mode shift in transportation. For this purpose, we need to measure basic metrics including mobility, cost and safety of taxi trips and the multimodal connectivity in urban environments. Third, the taxi mode has high accessibility and flexibility, since taxi services are not constrained to pre-specified origins and destinations in comparison with the users of public transit services. The broad sampling of OD trips enable us to study the collective behaviors of road users from their movements in the city, which is very valuable for elevating the comprehension of urban dynamics on the issues corresponding to system complexity such as the resilience of transportation systems and the relations between trip patterns and urban structures. In brief, a complete understanding on urban mobility and city dynamics is crucial for transforming cities to smart cities delivering effective, efficient, resilient, and sustainable services. Taxi data analysis provides us a prompt channel of information to gain insights of urban mobility and city dynamics, even though in the presence of many challenges on how to extract insights of urban mobility and city dynamics from the information and make them useful to stakeholders.

In this paper, we conduct a systematic analysis on taxi trip data to uncover urban mobility and city dynamics in multimodal transportation environments. We use a large-scale taxi origin-destination (O-D) trip data in Washington DC area as our main data source, and some other data from Open Data DC, the Maps API and the developer API as additional data sources. After an analysis on fundamental taxi trip characteristics, we apply data visualization, data analysis, statistical analysis and data fusion to investigate five important aspects from taxi trip data, where three concern urban mobility (respectively mobility and cost including effect of traffic congestion, trip safety, and multimodal connectivity), and two pertain to city dynamics (respectively transportation resilience, and the relation between trip patterns and land use). For each aspect, we uncover patterns and implications from taxi trip data, then explore the results to discuss qualitative and quantitative impacts of the inputs from key stakeholders on available measures of effectiveness on urban mobility and city dynamics, where key stakeholders include road users, system operators, and city planners and policymakers. At the end of the study, we summarize our findings to briefly discuss how to take advantages of the uncovered urban mobility and city dynamics to provide data-driven supports to key stakeholders in multimodal transportation environments.

\section{Data Description}

The DC Taxicab trip data contains the taxi trips for either pick up or drop off locations within District of Columbia (DC), which can be downloaded from Open Data DC provided by \citet{DC2017Data}. The data is provided by the Department of For-Hire Vehicles (DFHV) of DC, and does not include car-sharing vehicles such as Lyft or Uber. The information of each trip can be represented with a tuple $<l_o, l_d, t_o, t_{od}, d_{od}, c_{od}>$, where $l_o$ and $l_d$ are respectively the pickup and dropoff locations, $t_o$ and $t_{od}$ are respectively the pickup time (rounded to the nearest hour) and the whole trip time (in minutes), $d_{od}$ and $c_{od}$ are respectively the trip distance (in miles) and the total trip cost (in \$) including meter fare, tip, surcharge, tolls, and extras. There are no vehicle identifiers in the data set. We consider the data in a one-year period between [2015-09-01, 2016-08-31] for this study, which includes totally 14.34 million Taxicab trips.

We also use the following data sources: (1) The crash data of DC taxicabs between [2015-09-01, 2016-08-31] in Open Data DC for extracting taxi safety information; (2) The Maps API of \citet{GMap2017API} for extracting walk time information between locations; and (3) The developer API of \cite{FSQ2017Data} for extracting land use information.

Corresponding to spatial data processing and analysis, we use grid decomposition~\citep{castro2013taxi} in many cases to map geolocations into two dimensional grids for a better visualization or statistics. Specifically, we use the $k$-digit grid decomposition as shown in the Definition \ref{def:rDGD}.

\begin{definition} [\textbf{$k$-Digit Grid Decomposition}]
  \label{def:rDGD} 
    To transform geolocation to grid in spatial representation, the two values of each geolocation are respectively rounded to $k$ significant digits ($k \geq 0$) along latitude and longitude, i.e. the grid width and length are both $10^{-k}$ decimal degree along latitude and longitude.
\end{definition}

The larger $k$ is, the smaller each grid is (also the larger the number of grids is). In this paper, we consider $k \in \{2, 3\}$, which are respectively at the scales of neighborhoods and streets approximately.

\section{Basic Trip Characteristics}

We study basic characteristics of taxi trips by exploring their temporal and spatial distributions and by analyzing their probability distributions as functions of the key trip attributes such as trip distance, trip time, and trip cost.

\begin{figure} [h]
\centering

\begin{subfigure}{.49\textwidth} \centering \includegraphics[width=0.95\textwidth]{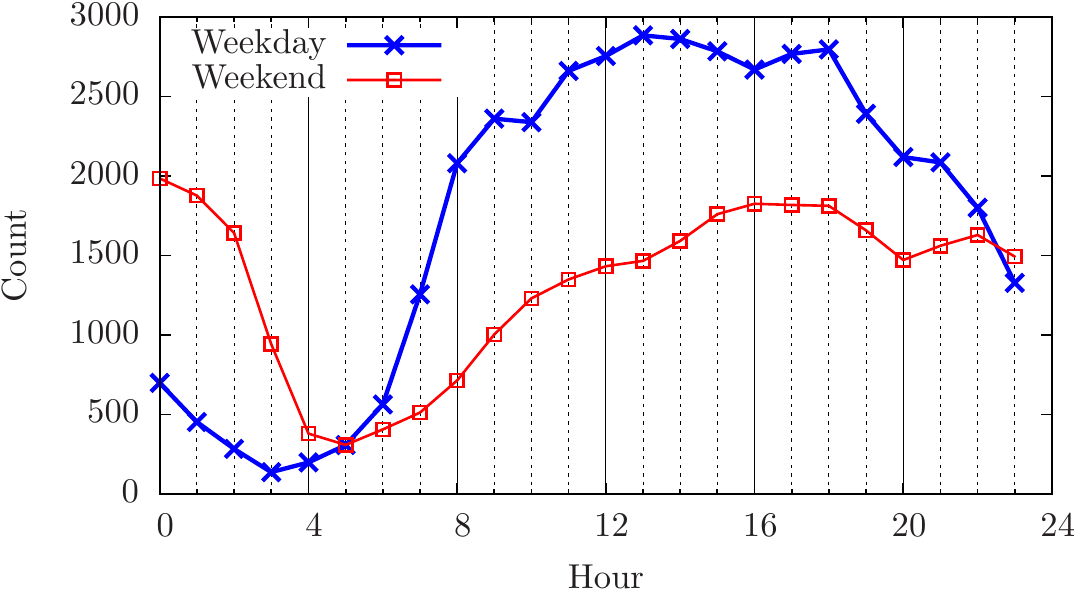} \caption{Trip Rates by Weekday and Weekend.}
\label{fig:dc_taxidata_1Y_hod_Week} 
\end{subfigure}
\begin{subfigure}{.49\textwidth} \centering \includegraphics[width=0.95\textwidth]{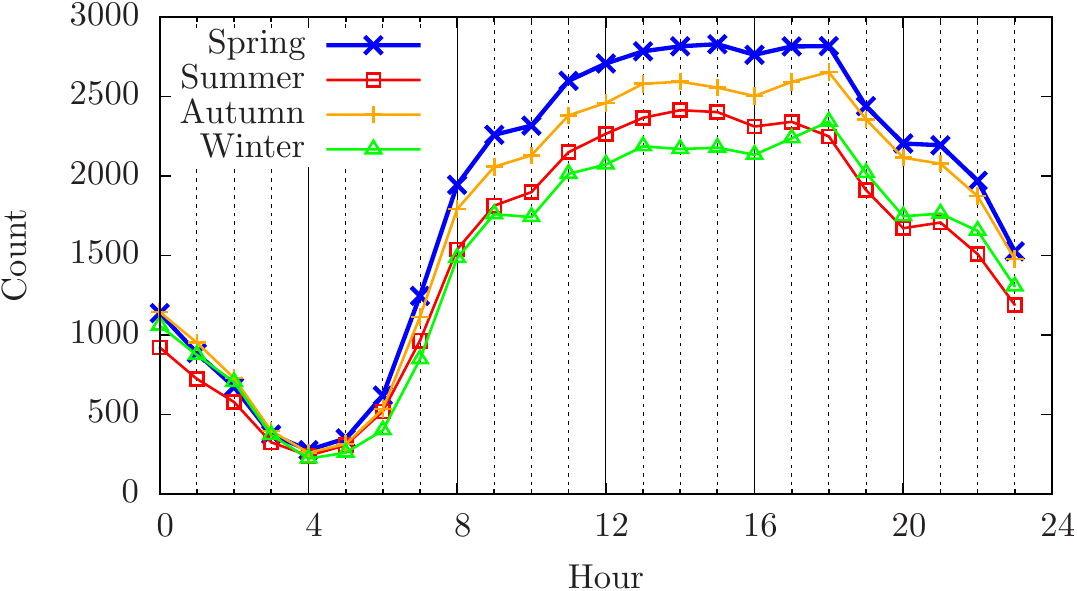} \caption{Trip Rates by Seasons.}
\label{fig:dc_taxidata_1Y_hod_Season} 
\end{subfigure}

\begin{subfigure}{.485\textwidth} \centering \includegraphics[width=0.95\textwidth]{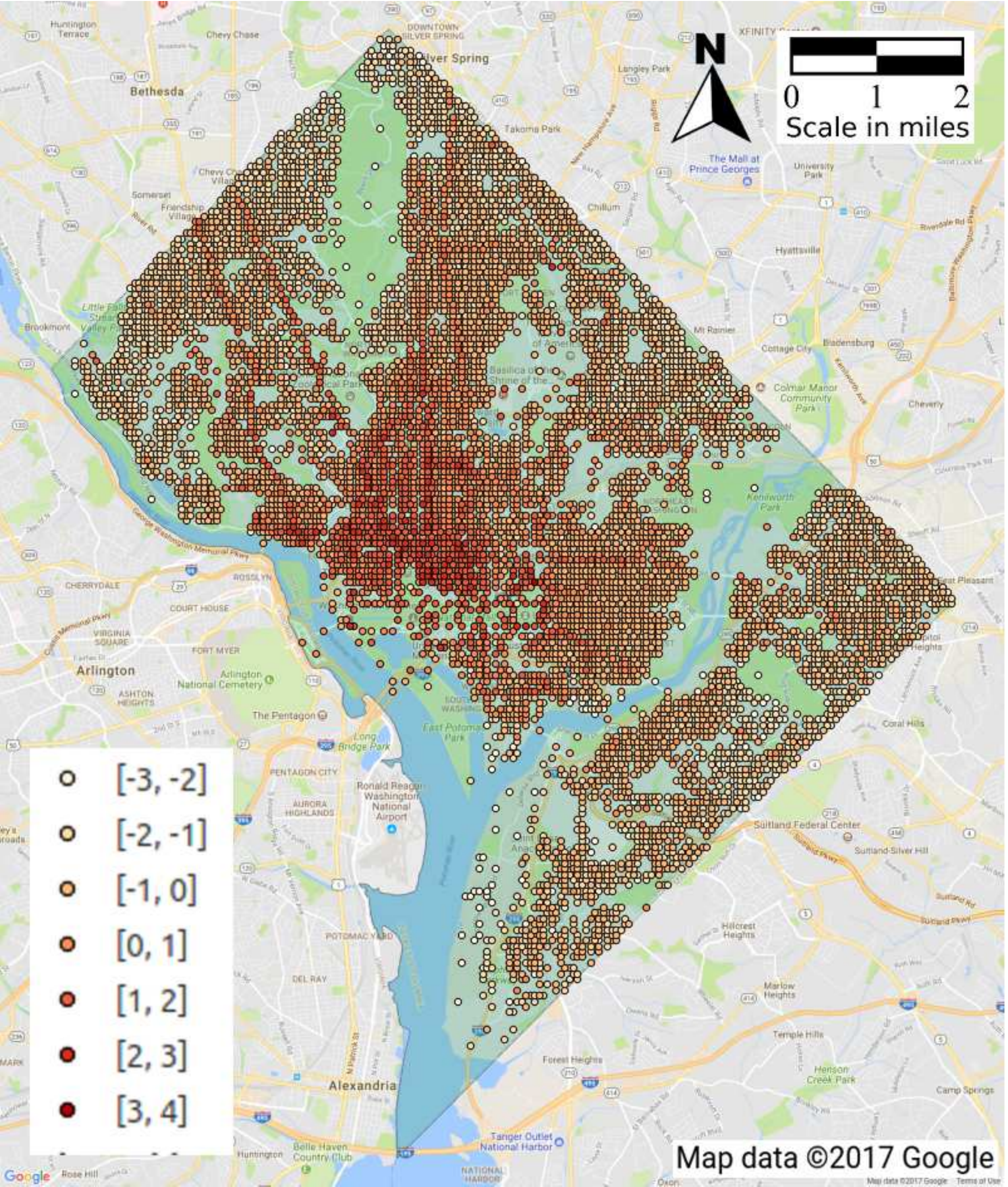} \caption{Spatial Trip Rates (in $\operatorname{log}_{10}$).}
\label{fig:dc_taxidata_1Y_geo_pick_r3} 
\end{subfigure}
\begin{subfigure}{.49\textwidth} \centering \includegraphics[width=0.95\textwidth]{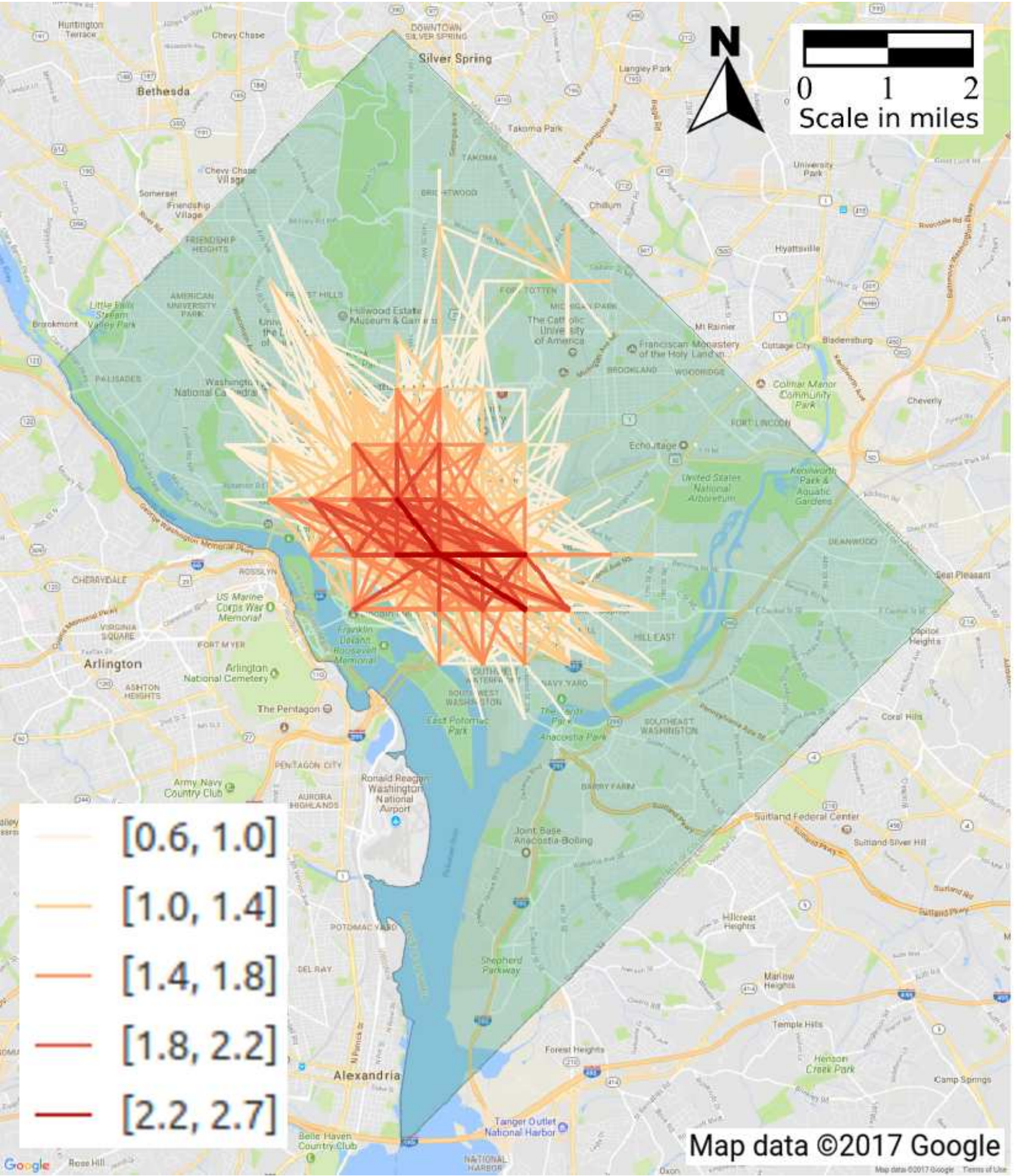} \caption{Spatial O-D Trip Rates (in $\operatorname{log}_{10}$).}
\label{fig:dc_taxidata_1Y_geo_od_r2_Sorted_1000} 
\end{subfigure}

\caption{Temporal and Spatial Distributions of Taxi Pickup Trips.}
\label{fig:dc_taxidata_1Y_spatial}
\end{figure}

Fig. \ref{fig:dc_taxidata_1Y_hod_Week} shows hourly taxi trip rates of both weekday (in blue) and weekend (in red), where each trip rate is computed with the number of trips averaged by its number of associated days. For weekday, although high taxi trip rate occurs during common working hours and the high rate continues even for a few after-work hours, no obvious AM or PM peak is expressed in the hour-of-day distribution of taxi trip rates, which indicates that commuting may not be the primary function of taxi trips. For weekend, a high value of taxi trip rate is shown during midnight. This is consistent with our common sense that taxi trip may play a significant role for night life of weekend. Fig. \ref{fig:dc_taxidata_1Y_hod_Season} gives the comparison of hourly taxi trip rates among different seasons, which clearly shows some seasonal variations in the taxi trip rates.  

Fig.~\ref{fig:dc_taxidata_1Y_geo_pick_r3} gives the spatial distribution of taxi trip rates by pickup locations, where pickup locations are decomposed with $k$-digit grids using $k=3$ (see Definition \ref{def:rDGD}). In Fig.~\ref{fig:dc_taxidata_1Y_geo_pick_r3}, each circle represents the center of a pickup location decomposed with $3$-digit grid, and its color represents the range of averaged daily taxi trip rate (i.e. averaged daily number of taxi trips) in $\operatorname{log}_{10}$ corresponding to the pickup location. As shown in the figure, the difference in the values of averaged daily taxi trip rate could reach even several orders of magnitude among different pickup locations. Fig.~\ref{fig:dc_taxidata_1Y_geo_od_r2_Sorted_1000} shows the spatial distribution of the top 1000 highest O-D pairs  in the ranking of taxi trip rates, where all O-D locations are decomposed with $k$-digit grids using $k=2$. In Fig.~\ref{fig:dc_taxidata_1Y_geo_od_r2_Sorted_1000}, each line represents an O-D pair, and its color represents the range of averaged daily taxi trip rate in $\operatorname{log}_{10}$ corresponding to the O-D pair. As shown in the figure, the top 1000 highest O-D pairs ranking in taxi trip rates are clustered around the central business district of DC.

\begin{figure} [h]
\centering

\begin{subfigure}{.49\textwidth} \centering \includegraphics[width=0.95\textwidth]{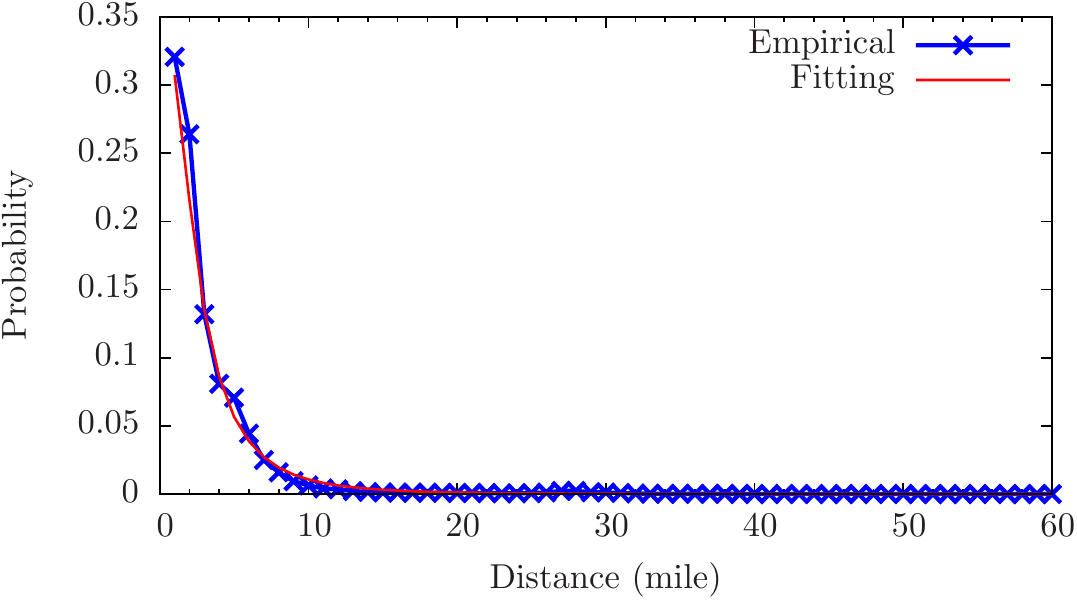} \caption{Empirical PDF of Trip Distance.}
\label{fig:dc_taxidata_1Y_grp_TripMileage_Sub_PDF} 
\end{subfigure}
\begin{subfigure}{.49\textwidth} \centering \includegraphics[width=0.95\textwidth]{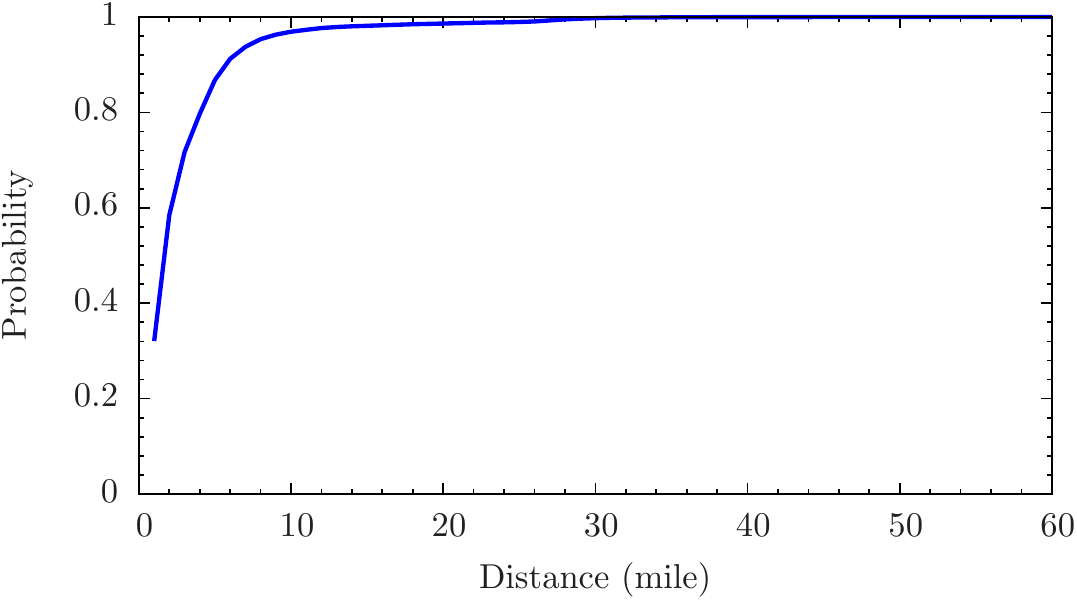} \caption{Empirical CDF of Trip Distance.}
\label{fig:dc_taxidata_1Y_grp_TripMileage_Sub_PDF_CDF} 
\end{subfigure}

\begin{subfigure}{.49\textwidth} \centering \includegraphics[width=0.95\textwidth]{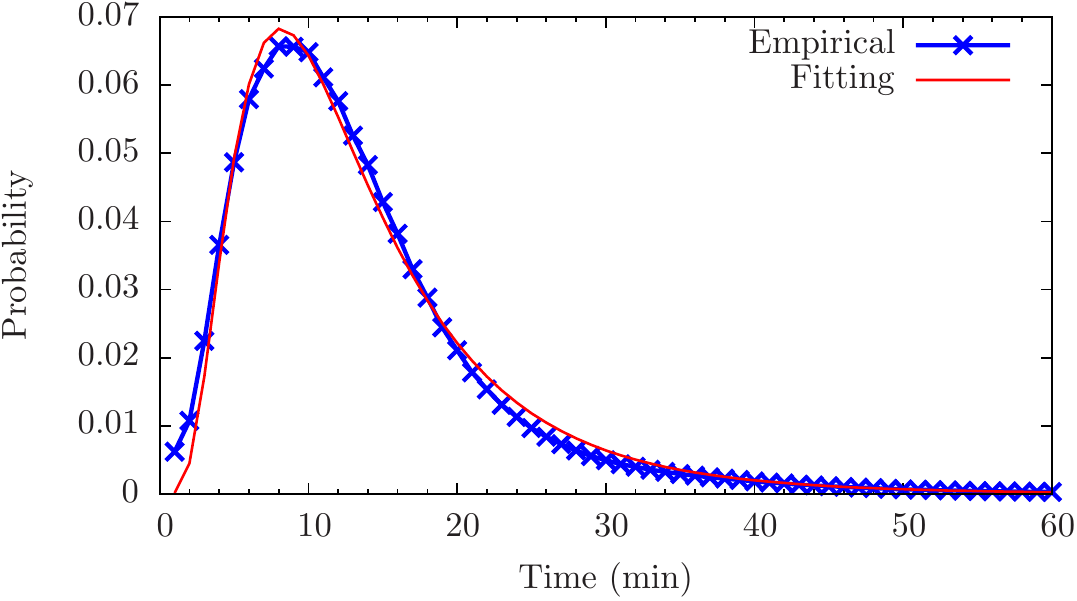} \caption{Empirical PDF of Trip Time.}
\label{fig:dc_taxidata_1Y_grp_TripTime_Sub_PDF_CDF} 
\end{subfigure}
\begin{subfigure}{.49\textwidth} \centering \includegraphics[width=0.95\textwidth]{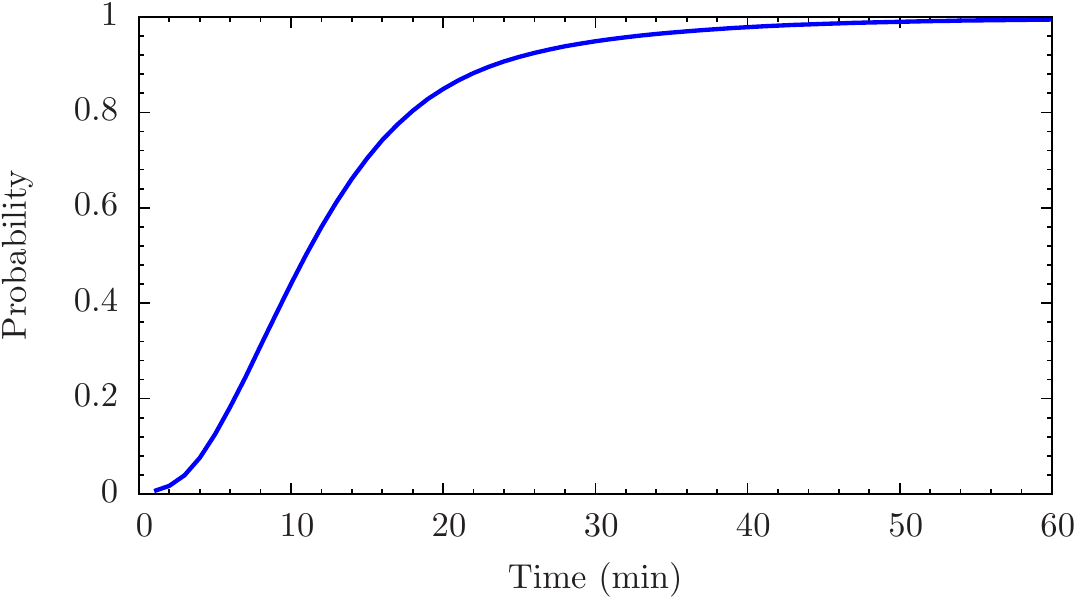} \caption{Empirical CDF of Trip Time.}
\label{fig:dc_taxidata_1Y_grp_TripTime_Sub_PDF} 
\end{subfigure}

\begin{subfigure}{.49\textwidth} \centering \includegraphics[width=0.95\textwidth]{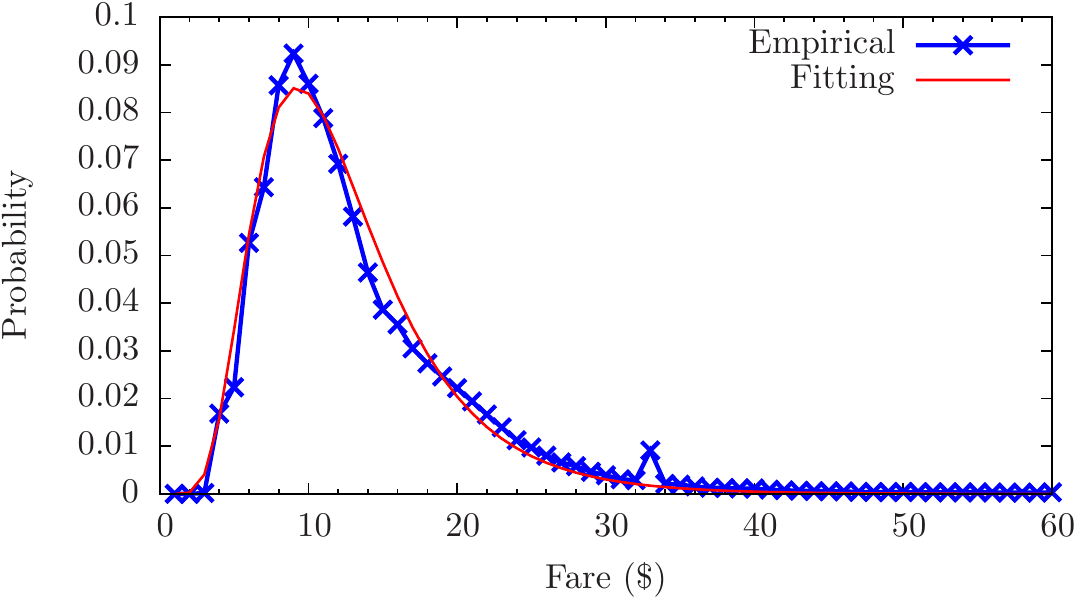} \caption{Empirical PDF of Trip Cost.}
\label{fig:dc_taxidata_1Y_grp_TotalCost_Sub_PDF} 
\end{subfigure}
\begin{subfigure}{.49\textwidth} \centering \includegraphics[width=0.95\textwidth]{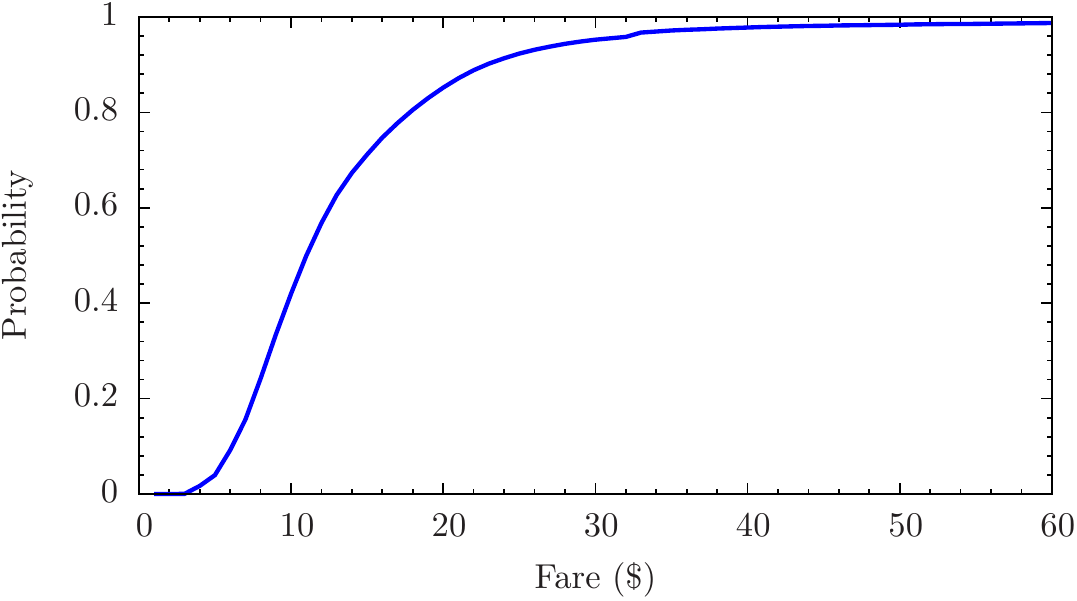} \caption{Empirical CDF of Trip Cost.}
\label{fig:dc_taxidata_1Y_grp_TotalCost_Sub_PDF_CDF} 
\end{subfigure}

\caption{Statistics of Distance, Time, and Cost of Taxi Trips.}
\label{fig:dc_taxidata_1Y_trip_stats_hod_Week}
\end{figure}

Fig. \ref{fig:dc_taxidata_1Y_trip_stats_hod_Week} shows the empirical Probability Density Functions (PDFs) and Cumulative Distribution Functions (CDFs) of distance, time, and cost of taxi trips. In the statistical analysis, we exclude the outliers with $t_{od}\le 0$, $d_{od}\le 0$ and $c_{od}\le 0$. 
We find that each of the three empirical PDFs can be well fitted with a lognormal distribution as shown below,
\begin{equation}
F(x | \mu, \sigma)=(x\sigma\sqrt{2\pi})^{-1}\operatorname{exp}({-\frac{\left(\ln x-\mu\right)^2}{2\sigma^2}}),
\label{eq:lognormal_distribution}
\end{equation}
where $x$ is a key trip attribute such as trip distance, time, and cost, and the parameters $\mu$ and $\sigma$ are estimated using a global optimization algorithm \citep{xie2014cooperative} minimizing the least squares between the data and the fitting function. For the three empirical PDFs, the parameters ($\mu$, $\sigma$) are respectively (0.7623, 0.9223), (2.4555, 0.6004), and (2.4326, 0.4564), and the root-mean-square errors (RMSE) are respectively 1.51E-3, 6.75E-4, and 1.54E-3. 
The trip distance distribution gives the median and the 90th percentile of taxi trip distance as 1.68 and 5.72 miles, respectively. 
The trip time distribution gives the median and the 90th percentile of taxi trip time as 10.96 and 23.40 minutes, respectively. 
The trip fare distribution gives the median and the 90th percentile of taxi trip cost as \$11.01 and \$22.85, respectively.

The two-parameter models on trip distance and time that are extracted from the taxi data in the DC area are consistent and complementary with the other recent works of the taxi data analysis in a few different cities~\citep{tang2015uncovering, xu2017mobility}. The lognormal distributions were found the best fittings to the trip distance and time of taxi services respectively by \cite{tang2015uncovering} for Harbin City and by \cite{xu2017mobility} for the New York City and two other cities of China. 
Moreover, our study shows that the trip cost of taxi services also follows the same parametric model. 
From the viewpoint of multimodal trip planning, the information of trip distance, time and cost is crucial for road users to make choice decision on transportation modes and travel routes. The two-parameter model is also useful for simulation studies on providing inputs. In addition, the model can be used for delivering a robust estimate of the mean value as $\operatorname{exp}(\mu+0.5 \cdot \sigma^2)$ in case that there exist significant outliers in data.

\section{Mobility and Cost: Effect of Traffic Congestion} \label{sec:mobcost}

\begin{figure} [h]
\centering

\begin{subfigure}{.49\textwidth} \centering \includegraphics[width=0.95\textwidth]{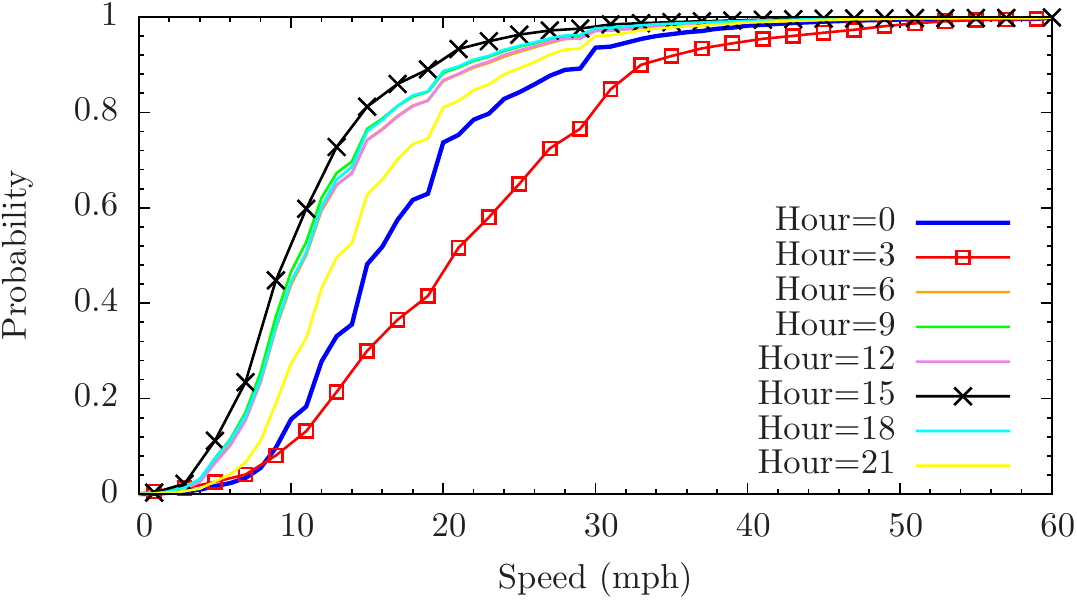} \caption{Taxi Trip Speed Distribution by ToDs.}
\label{fig:dc_taxidata_1Y_grp_TripSpeed_Weekday_+3_PDF_CDF} 
\end{subfigure}
\begin{subfigure}{.49\textwidth} \centering \includegraphics[width=0.95\textwidth]{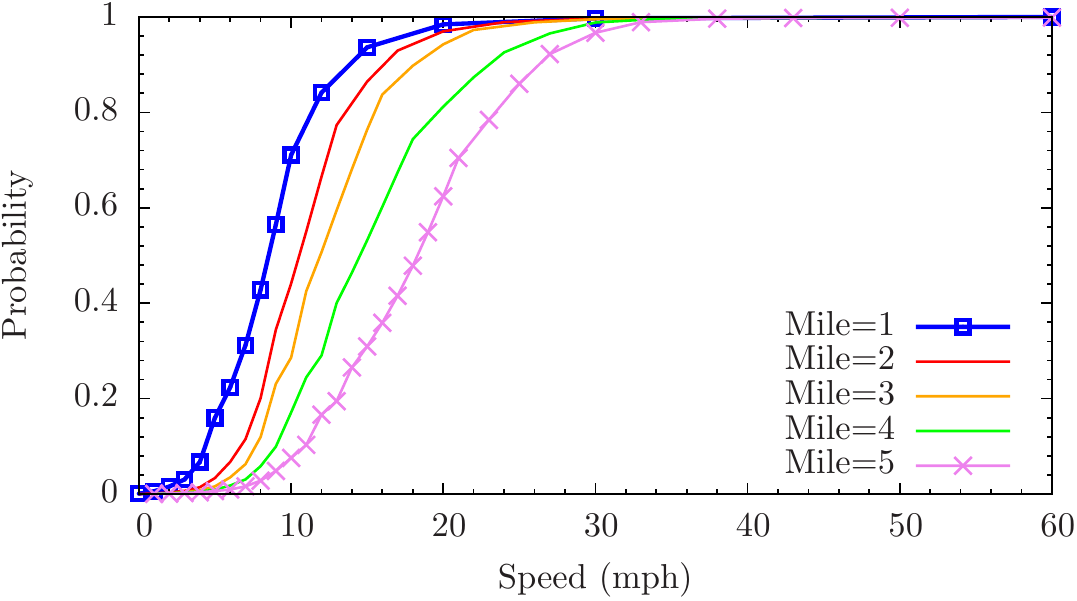} \caption{Taxi Trip Speed Distribution by Distances.}
\label{fig:dc_taxidata_1Y_grp_TripSpeed_miles_PDF_CDF} 
\end{subfigure}

\begin{subfigure}{.49\textwidth} \centering \includegraphics[width=0.95\textwidth]{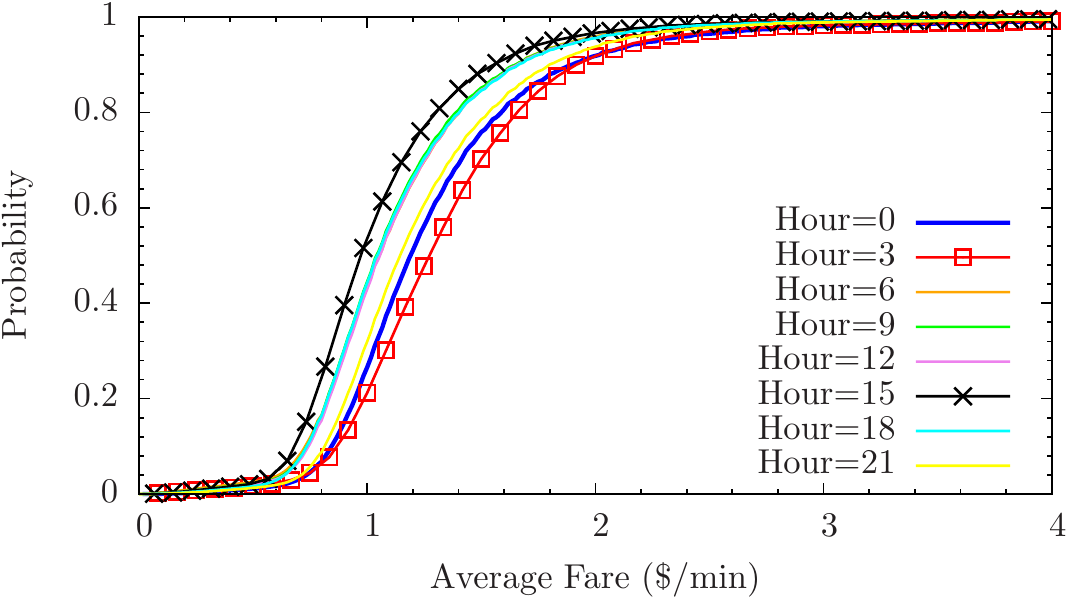} \caption{Taxi Trip Cost Distribution by ToDs.}
\label{fig:dc_taxidata_1Y_grp_TripPCost_+3_PDF_CDF} 
\end{subfigure}
\begin{subfigure}{.49\textwidth} \centering \includegraphics[width=0.95\textwidth]{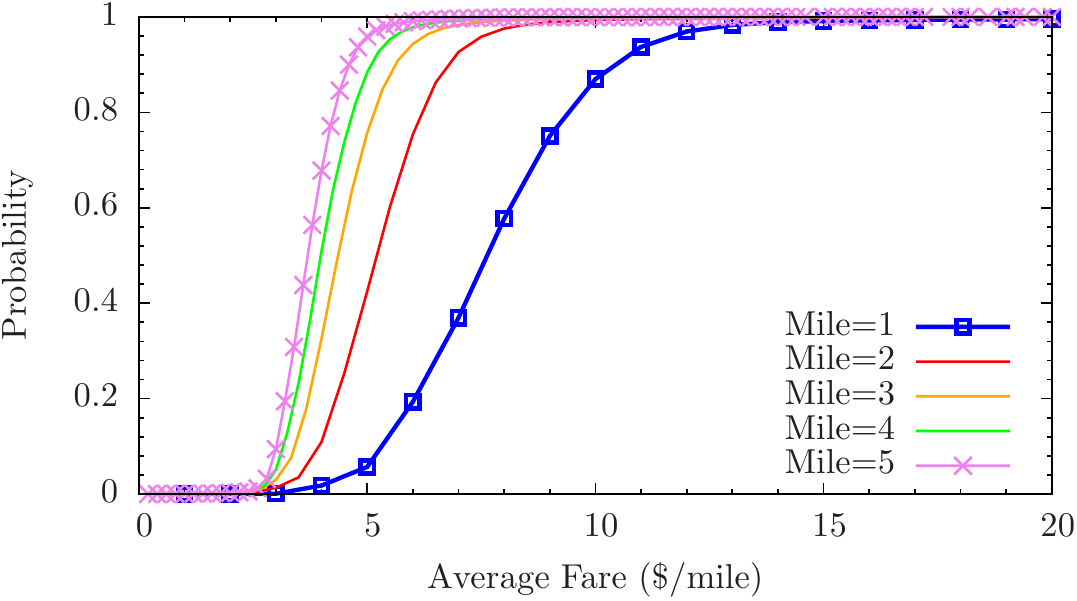} \caption{Taxi Trip Cost Distribution by Distances.}
\label{fig:dc_taxidata_1Y_grp_TotalCost_miles_PDF_CDF} 
\end{subfigure}

\caption{Distributions of Speed and Cost of Taxi Trips by ToDs and Distances.}
\label{fig:dc_taxidata_1Y_trip_speed_cost_CDF}
\end{figure}

Mobility and cost are two key factors to consider as people plan their travels. Fig. \ref{fig:dc_taxidata_1Y_trip_speed_cost_CDF} shows the comparisons of empirical CDFs of taxi trip speed (in mph) and cost (in \$/min and \$/mile) among different Times of Day (ToDs) and among different distances respectively, where the CDFs by ToDs are averaged over every three hours of day, and the CDFs by distance are averaged over every mile within five miles. For each panel of Fig. \ref{fig:dc_taxidata_1Y_trip_speed_cost_CDF}, we apply two-sample Kolmogorov-Smirnov tests on each pair of empirical CDFs to check if their difference in distributions is statistically significant. All the test results show that $p$-values approach to $0$, turning out that in each panel of Fig. \ref{fig:dc_taxidata_1Y_trip_speed_cost_CDF}, any two CDFs are different in statistical significance.

For each taxi trip $i$, the average speed is calculated as $v_{od}^{i}=d_{od}^{i}/t_{od}^{i}$. Fig. \ref{fig:dc_taxidata_1Y_grp_TripSpeed_Weekday_+3_PDF_CDF} shows the empirical CDFs of taxi trip speed at different ToDs. During 3-6 AM, the taxi trip speed is the highest, where the median and 90th percentile of speed are 18.16 and 32.01 mph, respectively (see the rightmost curve in Fig.\ref{fig:dc_taxidata_1Y_grp_TripSpeed_Weekday_+3_PDF_CDF}). During 15-18 PM, the taxi trip speed is the lowest, where the median and 90th percentile of speed are 9.94 and 19.71 mph, respectively (see the leftmost curve in Fig.\ref{fig:dc_taxidata_1Y_grp_TripSpeed_Weekday_+3_PDF_CDF}). 

Taxicabs can be seen as ``floating cars'' in urban traffic flows. Thus, speed of taxi trips can vary largely between different ToDs due to distinct urban congestion situations in traffic network. For measuring the impact of urban congestion, the {\em taxi-based congestion index} (TCI) is defined as
\begin{equation}
\text{TCI}=\tilde{V}_{\text{ToD}}^{max}/\tilde{V}_{\text{ToD}},
\end{equation}
where $\tilde{V}_{\text{ToD}}^{max}$ and $\tilde{V}_{\text{ToD}}$ are respectively the maximum median speed of all ToDs and the median speed of a specific ToD. Here $\tilde{V}_{\text{ToD}}^{max}$ can be seen as the speed in free flow. During a specific ToD, it is more congested if its $\tilde{V}_{\text{ToD}}$ is lower, which leads to a higher TCI. For the DC area, the maximum TCI in a day is then $\text{TCI}^{max}$=1.83, as $\tilde{V}_{\text{ToD}}$ is the lowest during 15-18 PM. Monitoring and understanding traffic congestion conditions of a city is critical for better urban planning. In the Urban Congestion Reports (UCR) by \citet{FHWA2017Congestion}, the {\em Travel Time Index} (TTI) is defined as $\overline{TT}/TT_{O}$, where $\overline{TT}$ and $TT_{O}$ are respectively the average travel time and free-flow travel time. TCI provides an alternative way to measure congestion, without requiring extensive traffic flow data from roadway sensors or track data from probing vehicles. 

Fig. \ref{fig:dc_taxidata_1Y_grp_TripSpeed_miles_PDF_CDF} shows the empirical CDFs of taxi trip speed for every mile within five miles in trip distance. It is shown that short-distance trips are slower than long-distance trips. This is likely due to the facts that taxi cabs often take arterial or highway for long-distance trips while have to use congested urban streets for short-distance trips most time. The median speeds are respectively 8.53, 10.54, 11.91, 14.53, and 18.31 mph corresponding to the taxi trip distance from one mile to five miles. Both Fig. \ref{fig:dc_taxidata_1Y_grp_TripSpeed_Weekday_+3_PDF_CDF} and \ref{fig:dc_taxidata_1Y_grp_TripSpeed_miles_PDF_CDF} show that a majority of taxi trips have a rather low median speed. In urban areas, traffic congestion is often serious due to the high density of intersections. The inefficient urban mobility might be vastly optimized and improved in practice by adopting some smart urban traffic control systems \citep{Xie2012,Xie2015c2tr,Papageorgiou:2003p488}.

Fig. \ref{fig:dc_taxidata_1Y_grp_TripPCost_+3_PDF_CDF} shows the empirical CDFs of taxi trip fares at different ToDs. The average highest and lowest taxi fares are respectively corresponding to the time period of 3-6 AM (the rightmost curve in Fig. \ref{fig:dc_taxidata_1Y_grp_TripPCost_+3_PDF_CDF}, concerning the least congested ToD as shown in Fig.\ref{fig:dc_taxidata_1Y_grp_TripSpeed_Weekday_+3_PDF_CDF}) and 15-18 PM (the leftmost curve in Fig. \ref{fig:dc_taxidata_1Y_grp_TripPCost_+3_PDF_CDF}, concerning the most congested ToD as shown in Fig.\ref{fig:dc_taxidata_1Y_grp_TripSpeed_Weekday_+3_PDF_CDF}), and their medians are respectively \$1.27 and \$0.97 per minute. This result is consistent with the recent study showing that the congestion often reduces the income of taxi drivers \citep{yuan2017modeling}. 

Fig. \ref{fig:dc_taxidata_1Y_grp_TotalCost_miles_PDF_CDF} shows the CDFs of taxi trip fares for every mile within five miles in trip distance. As shown in the figure, the shorter the trip distance, the higher the fare rate (i.e. fare per mile) is. As trip distance increases, taxi fare rate drops and will finally converge to a cost limit. 
Thus taxi is often considered by road users as a convenient and fast service for medium-to-long distance trips as compared to other transportation modes. 

From a conventional viewpoint, taxi service in urban area seems to be in a dilemma --- it is difficult for road users to take a taxi during traffic peak time, whilst the increase of taxi services leads to more traffic congestion ~\citep{tang2016two}. However, from another more recent viewpoint of Mobility-as-a-Service (MaaS)~\citep{jittrapirom2017mobility}, taxi system (along with other for-hire vehicle and shared mobility systems) in fact greatly helps reducing urban congestion as well as VMT. This is because in multimodal transportation environments, taxi services can provide pooled rides \citep{Sun2018Potential} and connect with other modes to make trips taken in a more sustainable way. As measures of effectiveness (MoE) on urban mobility and trip costs, Fig. \ref{fig:dc_taxidata_1Y_trip_speed_cost_CDF} is valuable for key stakeholders of taxi services such as road users, taxi drivers and operators, and city's urban planners and policymakers, as it can help them making better decisions on corresponding activities, operations, and policies, such as whether and when to choose taxi as travel mode, when and how to operate taxi services, how to plan and regulate taxi services, and etc.

\section{Trip Safety}  \label{sec:tripsafety}

Although taxi drivers are often considered as more experienced than normal drivers, a quite large amount of traffic violations and road crashes are found regarding to taxi drivers \citep{wu2016discrepancy}. Taxi drivers were shown to have a higher probability of speeding at the onset of yellow phase and of red-light running (RLR) at intersections \citep{wu2016discrepancy}. In addition, taxi drivers were shown to more likely be severely injured than normal drivers \citep{boufous2009factors}. 

\begin{figure} [htb]
\centering

\begin{subfigure}{.49\textwidth} \centering \includegraphics[width=.95\linewidth]{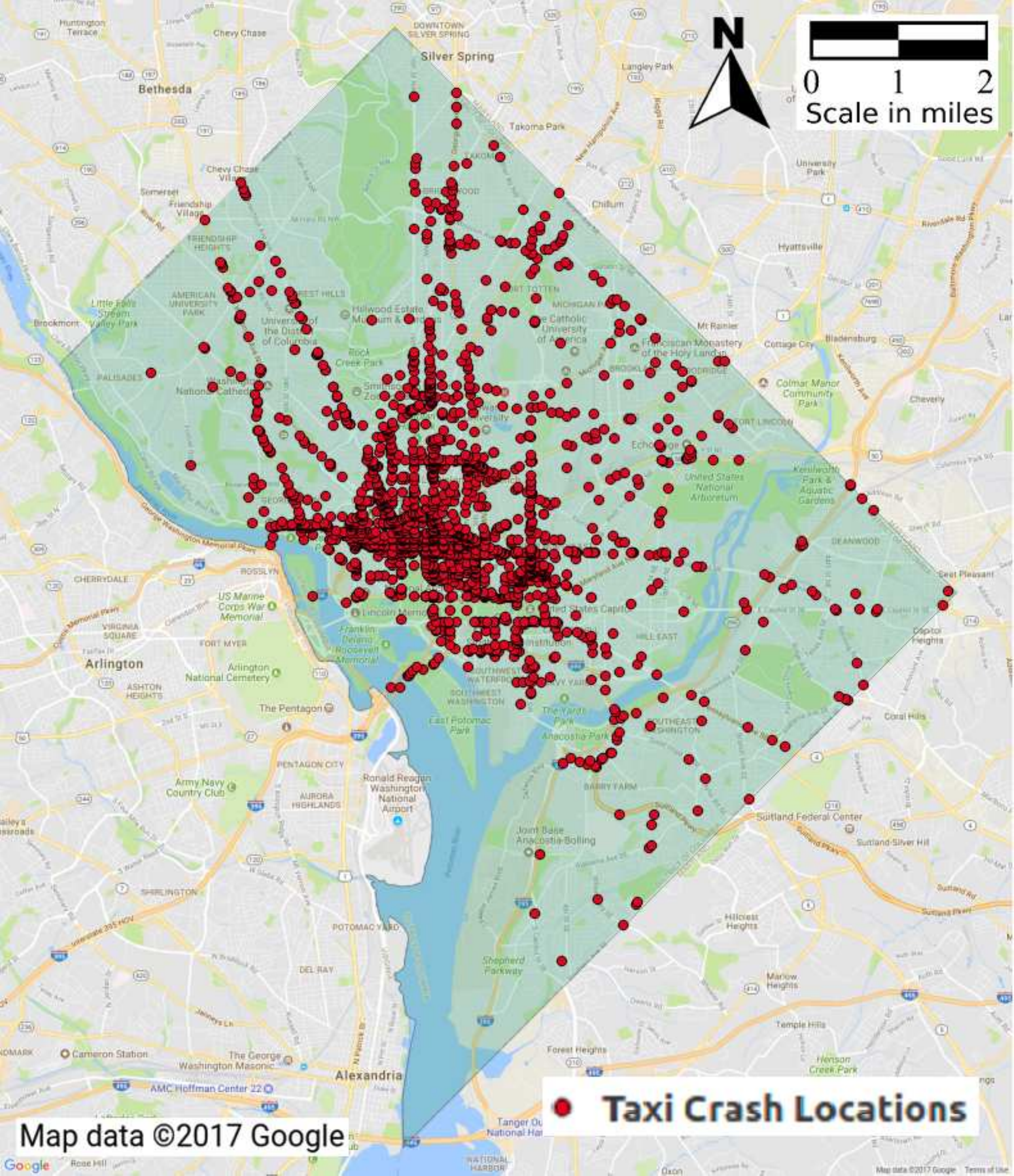} \caption{Taxi Crash Locations in DC.}
\label{fig:dc_crashes2_grp_Geo_TOTAL_TAXIS_Y2016} 
\end{subfigure}
\begin{subfigure}{.49\textwidth} \centering \includegraphics[width=.95\linewidth]{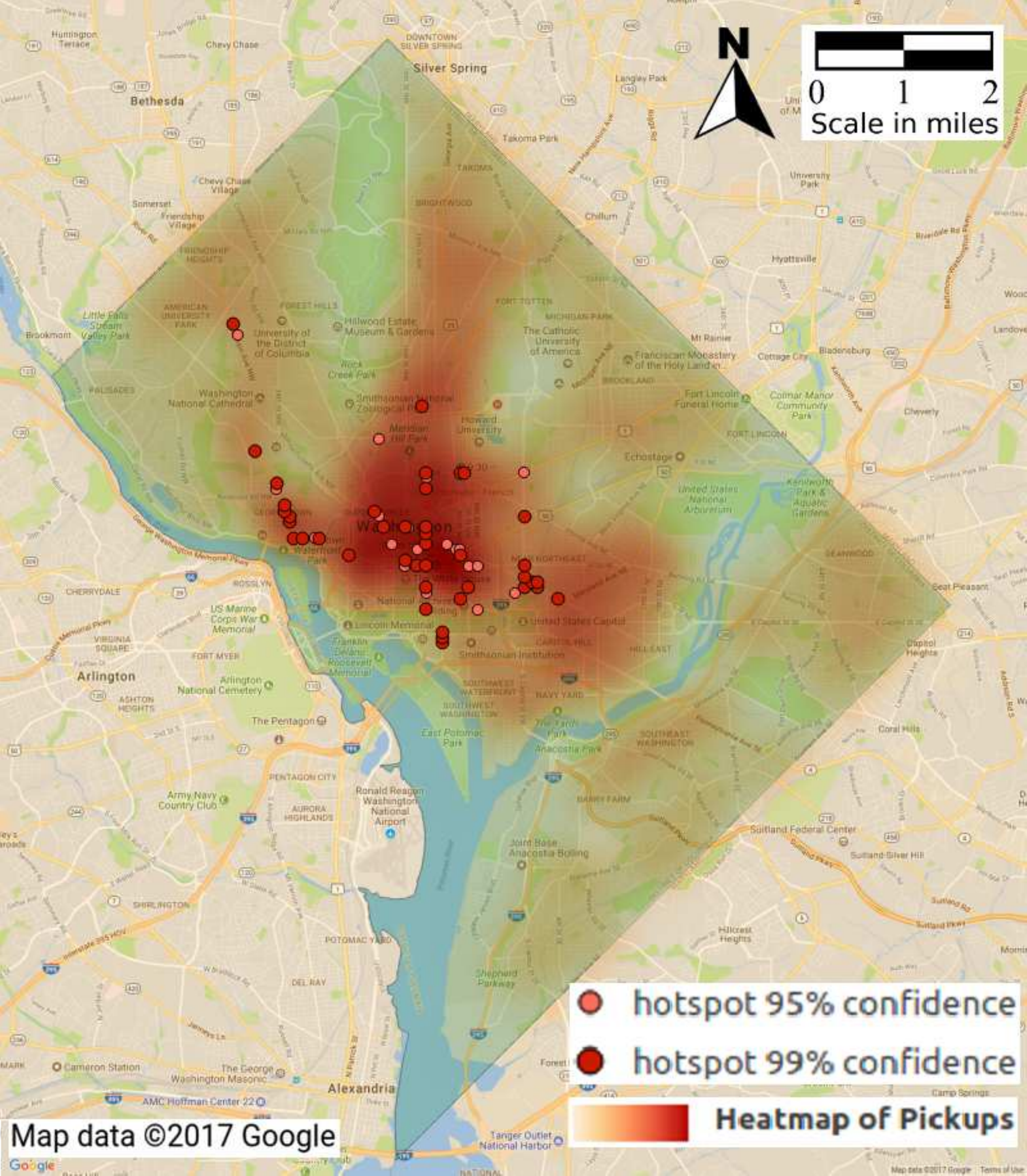} \caption{Crash Hotspots \& Heatmap of Taxi Pickups.}
\label{fig:dc_crashes2_grp_Geo_TOTAL_TAXIS_Y2016_PickupHeatmap_Hotspot} 
\end{subfigure}

\begin{subfigure}{.49\textwidth} \centering \includegraphics[width=0.95\textwidth]{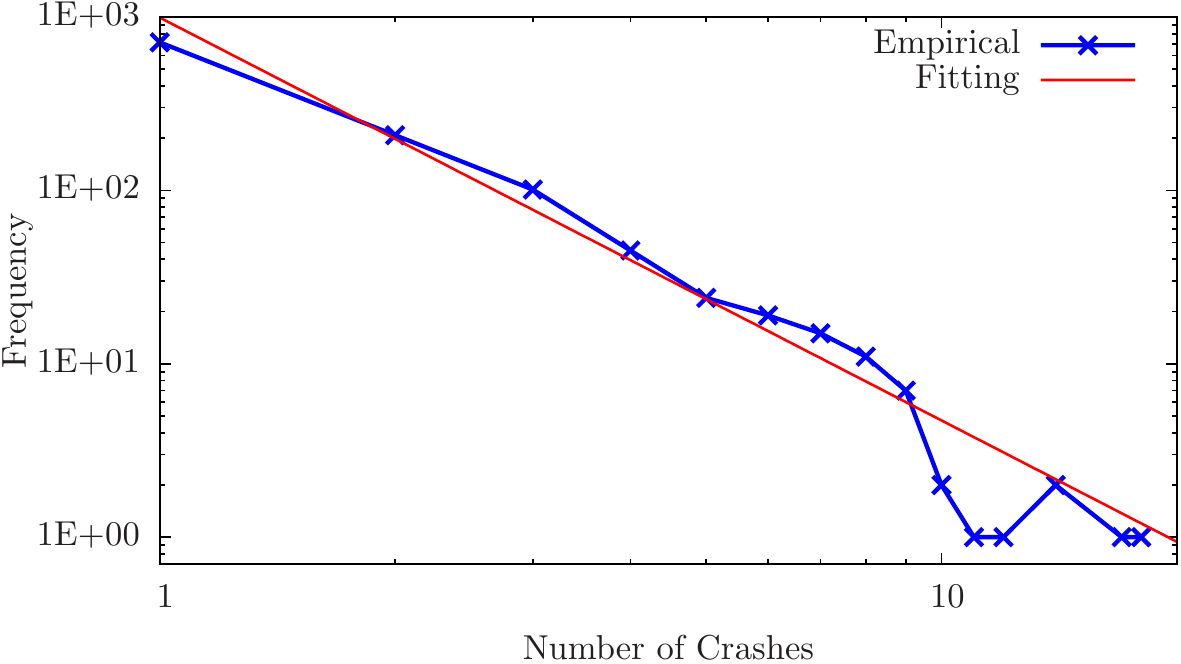} \caption{Frequencies of Crash Numbers.}
\label{fig:dc_crashes2_grp_Geo_TOTAL_TAXIS_1509_1Y_r3_hist} 
\end{subfigure}
\begin{subfigure}{.49\textwidth} \centering \includegraphics[width=0.95\textwidth]{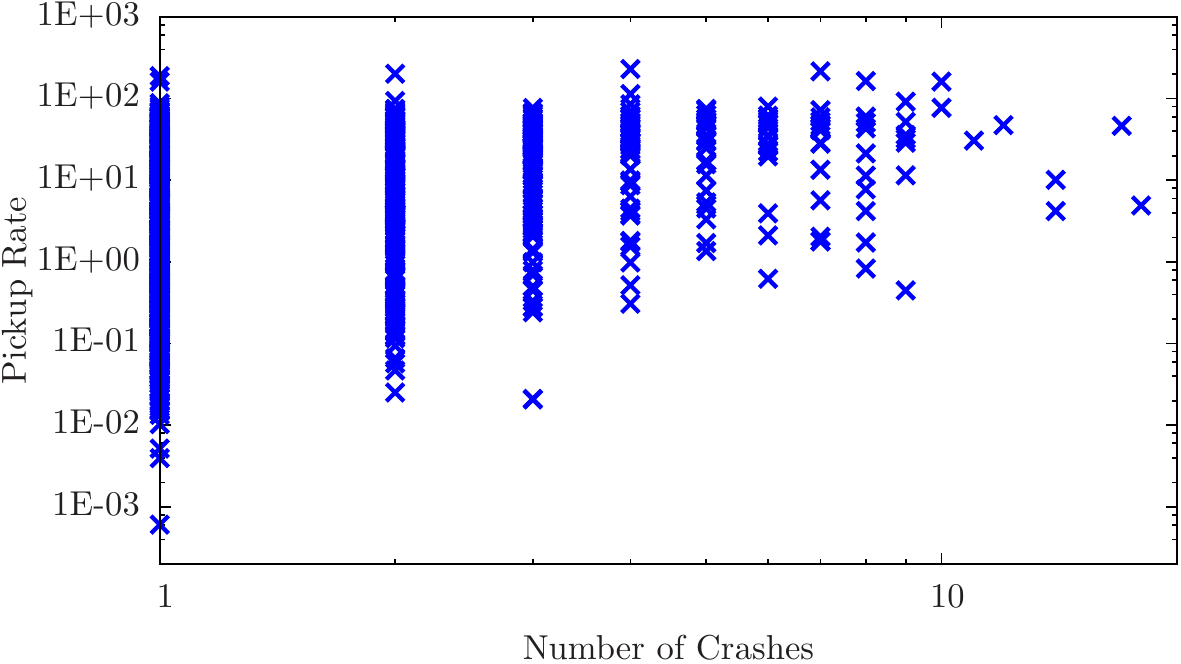} \caption{Crash Numbers vs. Taxi Demands.}
\label{fig:dc_crashes2_grp_Geo_TOTAL_TAXIS_1509_1Y_r3_n1_All} 
\end{subfigure}

\caption{DC Taxi-related Crashes and the Relevance with DC Taxi Demands.}
\label{fig:dc_taxidata_1Y_safety}
\end{figure}

According to the crash data in Open Data DC, there were 2,209 crashes involving taxicabs, among totally 28,730 vehicle crashes in DC during the studied period of [2015-09-01, 2016-08-31]. As shown in Fig. \ref{fig:dc_crashes2_grp_Geo_TOTAL_TAXIS_Y2016}, the spatial distribution is rather wide on the locations corresponding to taxi-related traffic crashes. To better understand taxi-related traffic crash probability of each location in DC, we perform a statistical analysis on the crash data. For each taxi-related traffic crash, we first decompose its location using 3-digit grid decomposition (see Def.\ref{def:rDGD}) to obtain its grid at the scale of streets approximately. Next, we group all crashes according to their grids and count crash number of each grid. Finally, for each crash number, we sum up all the corresponding grids to count the total number of grids for the crash number, i.e. the frequency of the crash number. Fig. \ref{fig:dc_crashes2_grp_Geo_TOTAL_TAXIS_1509_1Y_r3_hist} gives the frequencies of crash numbers in DC between [2015-09-01, 2016-08-31]. Notice that crash probability can be calculated from frequencies of crash numbers through a simple normalization over the total number of grids. As shown in Fig. \ref{fig:dc_crashes2_grp_Geo_TOTAL_TAXIS_1509_1Y_r3_hist}, the data can be fitted to a straight line in a log-log model with two coefficients $\beta_0$ and $\beta_1$, i.e., 
\begin{equation}
\operatorname{log} (\text{Frequency})=\beta_0+\beta_1 \operatorname{log} (\text{Number of Crashes}),
\label{eq:Freq_NCrashes}
\end{equation}
meaning that crash probability follows a common pattern of the scale-free law in the physical and social sciences~\citep{Xie2015,gonzalez2008understanding}. In one side, the high-accident locations (associated with large numbers of crashes in Fig. \ref{fig:dc_crashes2_grp_Geo_TOTAL_TAXIS_1509_1Y_r3_hist}) are subject to a small portion of streets. This means that if we could focus on the small group of high-accident streets and could put more safety precautions and endeavors such as enforcements, educations and engineering efforts on them to reduce crashes, transportation safety would be greatly improved. In another side, a large amount of streets are shown to have a small number of crashes, implying that the goal of decreasing crash number to zero would be challenging. 

Recent years, more and more cities have adopted Vision Zero \citep{tingvall2000vision,johansson2009vision}, a project aiming to achieve a highway system with no fatalities or serious injuries involving road traffic. It is affirmative to see more cities hold the vision that ``Life and health can never be exchanged for other benefits within the society' rather than the conventional comparison between costs and benefits on traffic safety, and the realization of Vision Zero in cities could be benefited from some innovative solutions, e.g., the smart Internet of things (IoT) \citep{xie2018key,Xie2018SIV} and big data analysis of traffic crashes \citep{xie2018multiscale}.

To further explore the high-crash locations,  we show the crash hot spots of taxi cabs in Fig. \ref{fig:dc_crashes2_grp_Geo_TOTAL_TAXIS_Y2016_PickupHeatmap_Hotspot}, where heatmap of the taxi pick-up flows from Fig. \ref{fig:dc_taxidata_1Y_geo_pick_r3} is included to observe if there is any correlation between traffic crash and trip demand. The statistically significant crash hot spots are extracted with the results of z-scores from the Gi* statistic by \citet{ord1995local} with 95\% and 99\% confidences respectively. The heatmap is computed using kernel density estimation. Fig. \ref{fig:dc_crashes2_grp_Geo_TOTAL_TAXIS_Y2016_PickupHeatmap_Hotspot} indicates that most of crash hot spots are located at the regions with a high trip demand. To further quantitatively study the correlation between taxi crash probability and taxi pickup demand, we combine the statistical analysis with 3-digit grid decomposition for Fig. \ref{fig:dc_crashes2_grp_Geo_TOTAL_TAXIS_1509_1Y_r3_hist} and Fig.~\ref{fig:dc_taxidata_1Y_geo_pick_r3}. For each grid, we can get its number of taxi crashes from the analysis for Fig. \ref{fig:dc_crashes2_grp_Geo_TOTAL_TAXIS_1509_1Y_r3_hist} and its taxi pickup demand from Fig.~\ref{fig:dc_taxidata_1Y_geo_pick_r3}. We estimate taxi pickup demand of each grid using the average pickup rates of its $({2\cdot n+1})^{2}$ nearest-neighbor grids with a cutoff range of $[-n, n]$ grids along both latitude and longitude. Here we use $n=1$ in this paper. Fig. \ref{fig:dc_crashes2_grp_Geo_TOTAL_TAXIS_1509_1Y_r3_n1_All} shows the relationship between the numbers of taxi crashes and the taxi pickup demands in 3-digit grids, i.e. at about the street scale. As shown in the figure, the growth in traffic demand may increase the probability of crashes occurring, but not always. Fig. \ref{fig:dc_crashes2_grp_Geo_TOTAL_TAXIS_1509_1Y_r3_n1_All} indicates that high traffic demand is a necessary but not a sufficient condition causing a high probability of crashes. It would be valuable to study the features of the low-crash zones with high traffic demand and apply the obtained knowledge for reducing crashes.  
 
\section{Multimodal Connectivity} \label{sec:connectivity}

\begin{figure} [h]
\centering

\begin{subfigure}{.49\textwidth} \centering \includegraphics[width=0.95\textwidth]{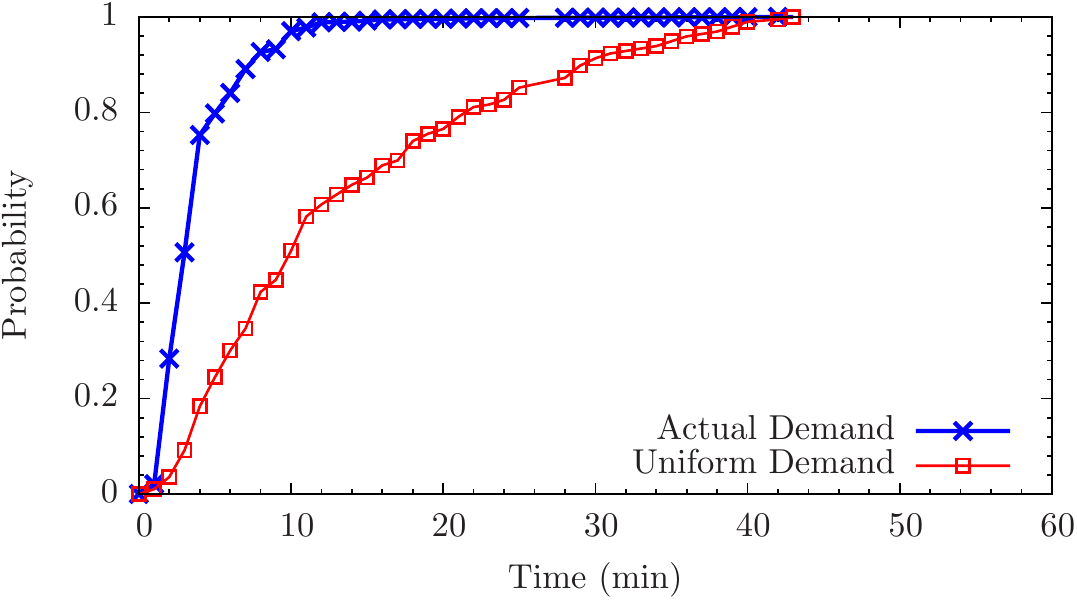} \caption{Walk Time to Bikeshare Stations.}
\label{fig:TaxiG2OBikeD_GDM_Walk_min_osid_eTime_grp_Time_PDF_CDF}  
\end{subfigure}
\begin{subfigure}{.49\textwidth} \centering \includegraphics[width=0.95\textwidth]{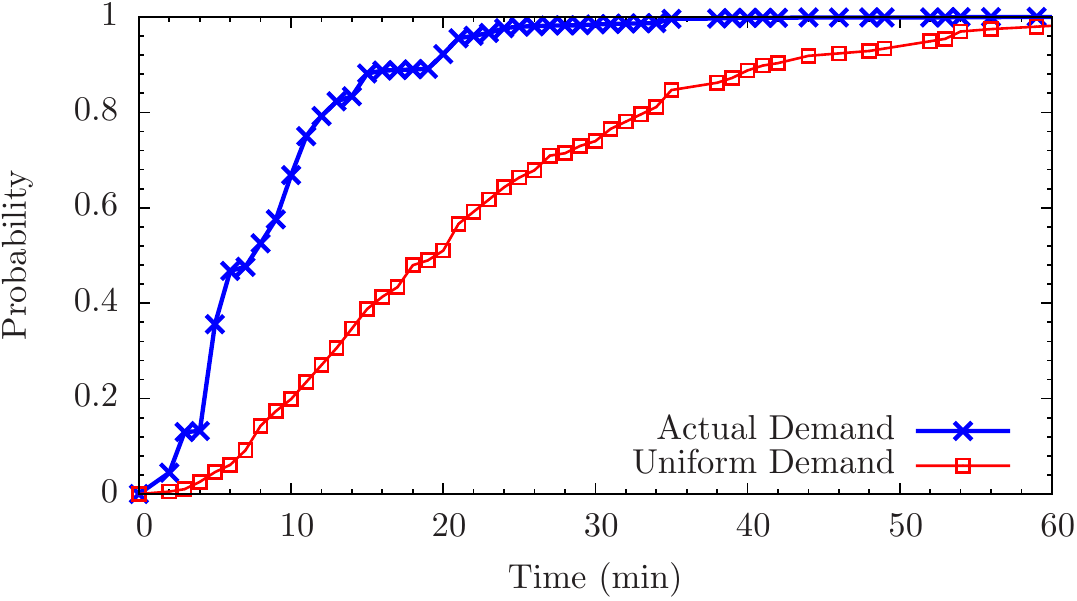} \caption{Walk Time to Metro Stations.}
\label{fig:TaxiG2OMetroD_GDM_Walk_min_osid_eTime_grp_Time_PDF_CDF} 
\end{subfigure}

\begin{subfigure}{.49\textwidth} \centering \includegraphics[width=0.95\textwidth]{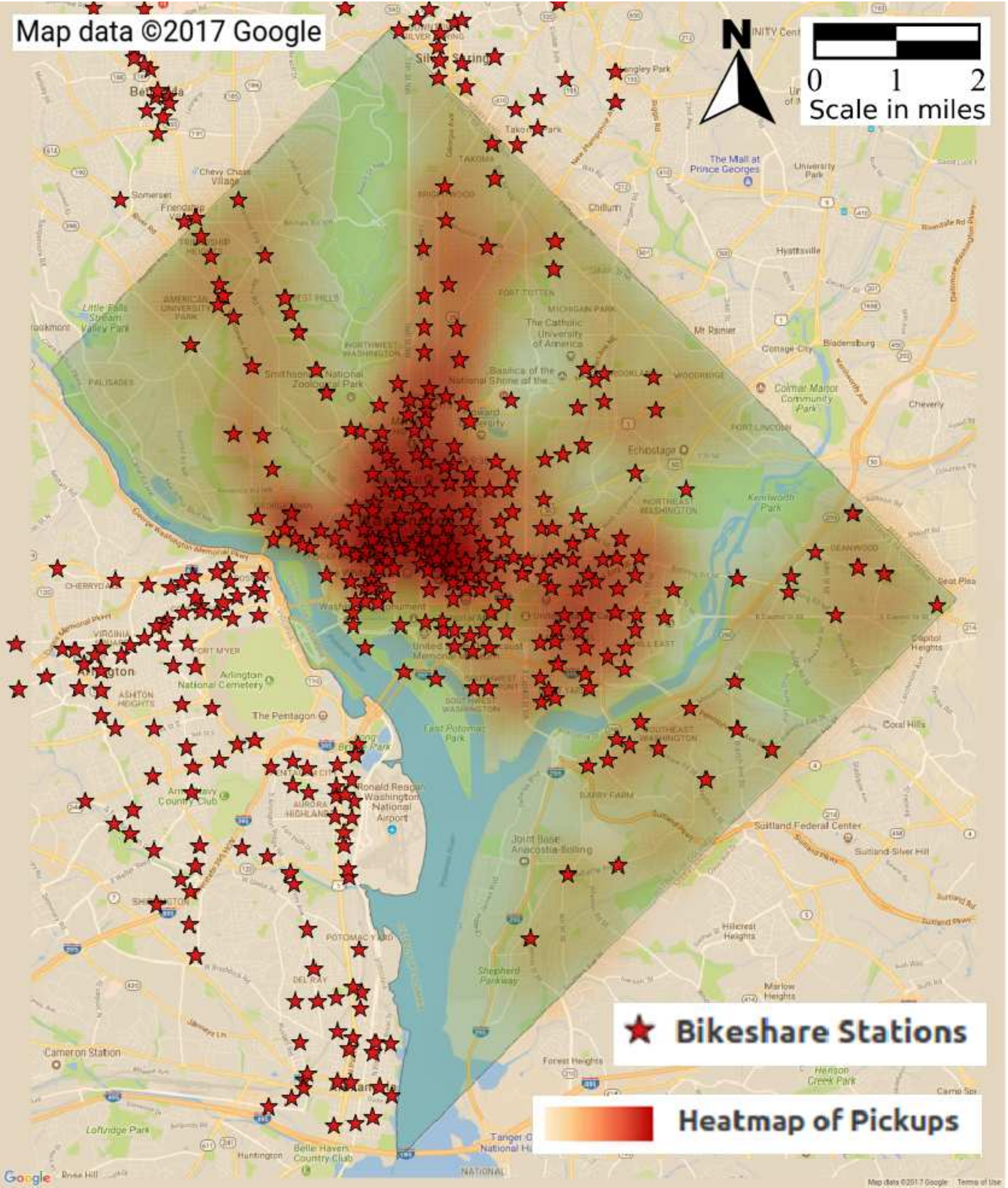} \caption{Taxi Pickup Heatmap \& Bikeshare Stations.}
\label{fig:dc_taxidata_1Y_geo_regions_r3_heatmap_bikestations}  
\end{subfigure}
\begin{subfigure}{.49\textwidth} \centering \includegraphics[width=0.95\textwidth]{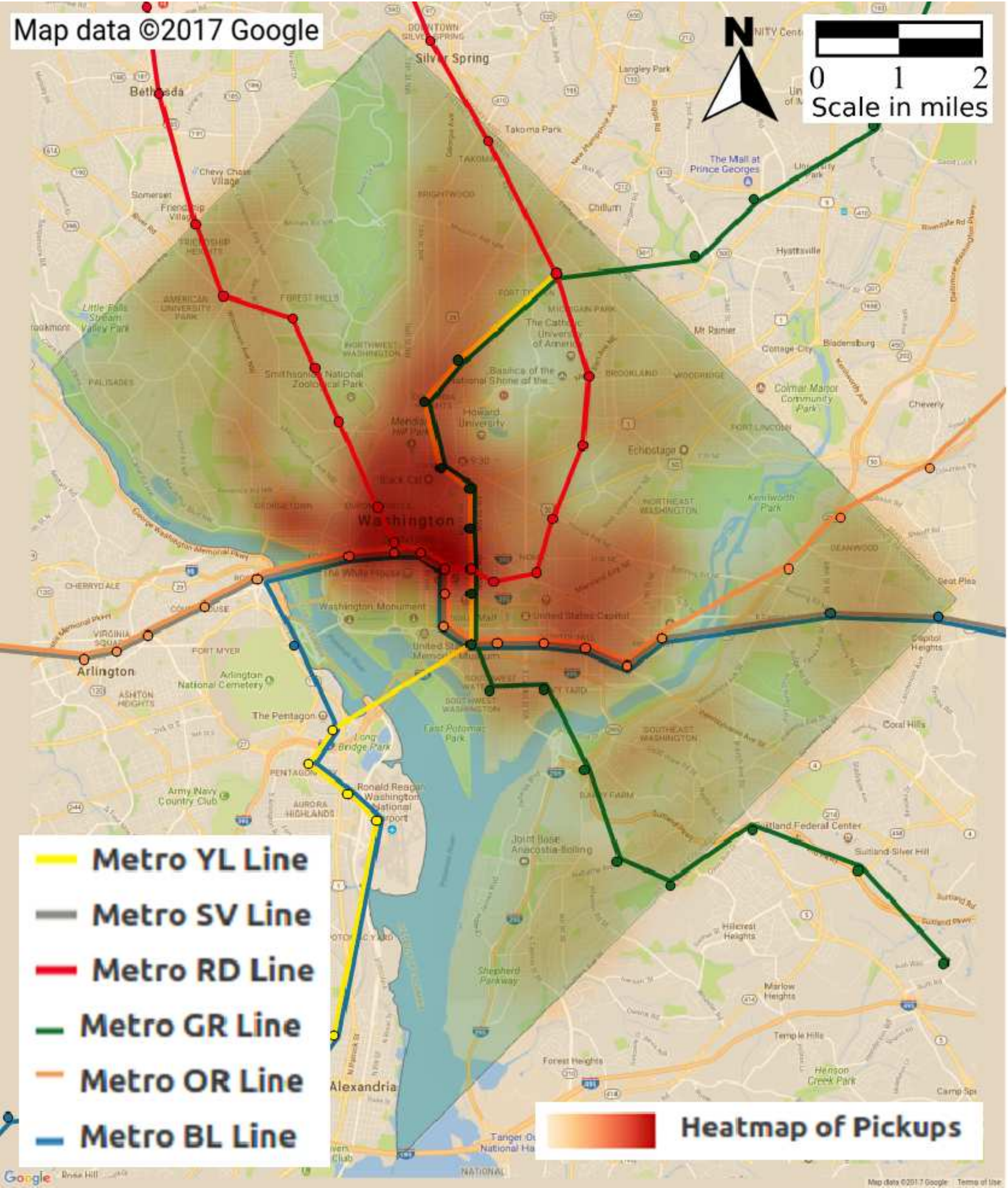} \caption{Taxi Pickup Heatmap \& Metro Stations.}
\label{fig:dc_taxidata_1Y_geo_regions_r3_heatmap_metrostations} 
\end{subfigure}

\caption{Walk-Time Distributions from Taxi Pickup Locations to Bikshare or Metro Stations and Spatial Distributions of Bikshare and Metro Stations \& Taxi Pickup Heatmap.}
\label{fig:dc_taxidata_1Y_multimodal_connections}
\end{figure}

Delivering excellent connectivities between different traffic modes is a key to maintain the efficiency of a multimodal transportation system. Fig. \ref{fig:dc_taxidata_1Y_multimodal_connections} shows the walk-time and spatial distributions from any taxi pickup location to its nearest Capital Bikeshare or nearest Metro stations, where the walk time is retrieved using Google Maps API. The probability of the walk time from a taxi pickup location to its nearest station is computed with the method as shown below:
\begin{enumerate}
\setlength{\itemsep}{-1pt}
\item Transform geolocations to two dimensional grids with $k$-digit grid decomposition (see Def.\ref{def:rDGD}), where $k=2$ (i.e., at about the neighborhood scale) is used to decrease the data retrieving cost in using Google Maps API but retain sufficiently fine granularity.  
\item For each grid in map, retrieve its minimum walk time to the station among the $K$ nearest stations ($K=20$ by default) in the straight line distance from Google Maps API.
\item Let $M(t)$ be the set of grid indices, where for each grid $g_m$ with $m\in M(t)$, the minimal walk time from the grid to a station is $t$ (for which we use binned time in one-minute intervals). Let $C_{p}(g_m)$ be the number of taxi pickups (i.e. actual demand) in each grid $g_m$. Then the probability mass functions of the walk-time distributions respectively for uniform demand and for actual demand can be calculated as
\begin{equation}
F_{ud}(t) = \frac{|M(t)|}{\sum_t |M(t)|}, ~~~~ F_{ad}(t) = \frac{\sum_{m\in M(t)}C_{p}(g_m)}{\sum_t \sum_{m\in M(t)}C_{p}(g_m)}. 
\label{eq:walktimedist}
\end{equation}
\end{enumerate}

Figs. \ref{fig:TaxiG2OBikeD_GDM_Walk_min_osid_eTime_grp_Time_PDF_CDF} and \ref{fig:TaxiG2OMetroD_GDM_Walk_min_osid_eTime_grp_Time_PDF_CDF} give the comparisons of CDFs of the walk-time distributions between uniform demand (red) and actual demand (blue) on connectivity from taxi pickup locations to Capital Bikeshare and Metro stations respectively. Figs. \ref{fig:dc_taxidata_1Y_geo_regions_r3_heatmap_bikestations} and \ref{fig:dc_taxidata_1Y_geo_regions_r3_heatmap_metrostations} give the spatial distributions of Capital Bikeshare and Metro stations respectively in the DC area, where the actual demand on connectivity (i.e. the heatmap of taxi pickups) is also shown in both of the figures. 

As shown in Figs. \ref{fig:TaxiG2OBikeD_GDM_Walk_min_osid_eTime_grp_Time_PDF_CDF} and \ref{fig:TaxiG2OMetroD_GDM_Walk_min_osid_eTime_grp_Time_PDF_CDF}, all the walk-time CDFs of actual demand (blue) is on the top left of those of uniform demand (red). It proves that accessibility in connections from taxi to other transportation modes has a strong positive impact on taxi demand. More accessibility to connect other transportation modes a location has, more taxi demands it has. The data shows that 50\% and 90\% taxi passengers in DC can access the nearest bikeshare station in 2.97 and 7.26 minutes respectively (see Fig. \ref{fig:TaxiG2OBikeD_GDM_Walk_min_osid_eTime_grp_Time_PDF_CDF}), and can access the nearest metro station in 9.83 and 29.13 minutes respectively (see Fig. \ref{fig:TaxiG2OMetroD_GDM_Walk_min_osid_eTime_grp_Time_PDF_CDF}). We find that the distributions of the walk time from any taxi drop-off location to the nearest stations are similar to the results using taxi pickup locations. The analysis results are independent from the choice of using taxi pickups or drop-offs or both.

In a multimodal environment, accessibility in connections between different modes has significant impacts on users to select which transportation modes to use. The impacting factors on accessibility in connectivity include level of convenience, costs, speed, distance and etc. Transportation policy makers might have to consider all the impacting factors if expecting people use more sustainable transportation modes such as bike ride or public transits, rather than private cars. Let us take an example on replacing vehicle by bike. As shown in Fig.~\ref{fig:dc_taxidata_1Y_grp_TripSpeed_Weekday_+3_PDF_CDF}, taxi speed is rather low during rush hours in DC, so are the other vehicles. In this case, many road users would change their transportation mode to bike ride during rush hours given their trips are not long-distance. In fact, many taxi trips are short-distance demands in DC (see Fig.~\ref{fig:dc_taxidata_1Y_trip_stats_hod_Week}), therefore it is feasible for road users in DC to replace some of taxi trips with bike rides. Beyond improving sustainability of DC, this transportation mode shift could greatly reduce traffic congestion.

In most cases, it is not easy for road users to replace private cars directly with public transits. The comparison between Fig.~\ref{fig:TaxiG2OBikeD_GDM_Walk_min_osid_eTime_grp_Time_PDF_CDF} and Fig.\ref{fig:TaxiG2OMetroD_GDM_Walk_min_osid_eTime_grp_Time_PDF_CDF} shows that the walk time (reflecting the distance) for road users to access the nearest Metro stations is likely much longer than to bikeshare stations. Concerning connection with public transit, both taxi and bike trips are turned out to be very helpful on providing convenient intermediate or last-mile solutions~\citep{wang2017new}. However, the fare of taxi might be one barrier for some road users on choosing taxi as their connection mode to metro. As shown in Fig.~\ref{fig:dc_taxidata_1Y_grp_TotalCost_miles_PDF_CDF}, the fare of taxi for short-distance trip is much higher than that for long-distance. In this case, if these road users dislike biking either, it would be hard for them to choose metro as their transportation mode. Reduction in the fare of taxi might be realized through improving efficiency of the taxi system~\citep{zhan2016graph}, where either the balance between demand and supply could be optimized leading to less empty trips \citep{zhang2016impacts}, or the rate of ride sharing could be increased on each individual taxi trip \citep{tachet2017scaling}. 

\section{Transportation Resilience} \label{sec:resilience}

An urban transportation system might suffer various levels of disruptions or extreme events due to many reasons such as emergency maintenance and extreme weathers. Transportation resilience is defined to evaluate its responsive ability to a disruption or extreme event, where the static resilience often refers to a transportation system's capability of maintaining its basic function during a major perturbation, while the dynamic resilience refers to how quickly a transportation system can recover its state of normal function after a major perturbation~\citep{mattsson2015vulnerability,donovan2017empirically}.

On 03-16-2016 (Wednesday), the DC Metrorail system, which is the backbone of DC public transportation systems, was entirely shutdown for emergency repairs. The impact of this major disruption was investigated by a systematic research \citep{pu2017state}. The research revealed that vehicle volumes, bus ridership, and Capital Bikeshare ridership all increased in the regional core area, but did not study taxi and other ride-sharing systems. 

\begin{figure} [ht]
\centering

\begin{subfigure}{.49\textwidth} \centering \includegraphics[width=.95\linewidth]{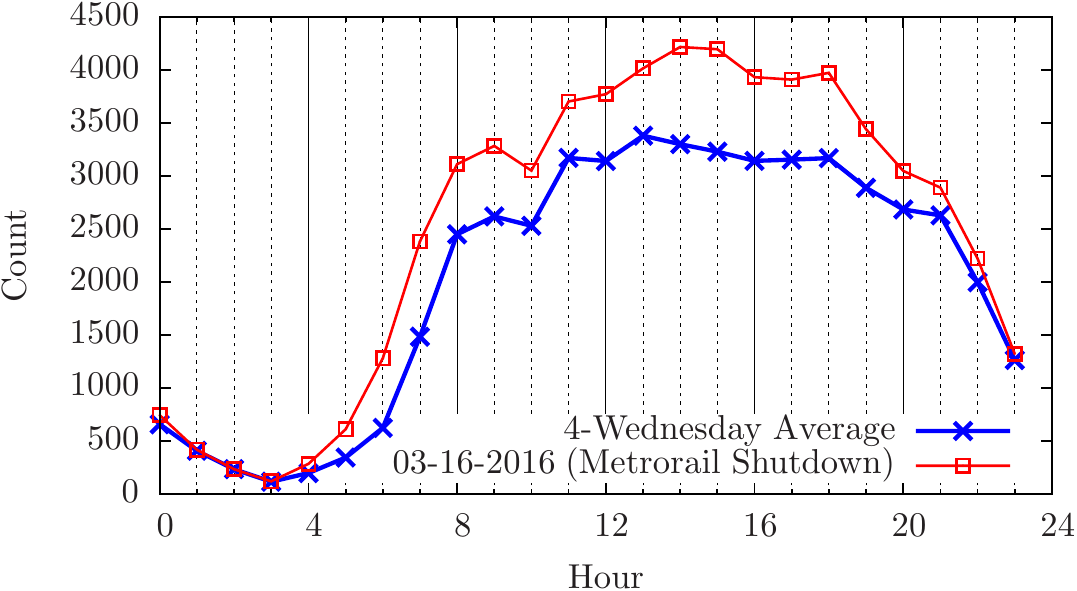} \caption{Changes in Taxi Trip Volumes.}
\label{fig:dc_taxidata_Metro16RangeVsDate_hod_Week} 
\end{subfigure}
\begin{subfigure}{.49\textwidth} \centering \includegraphics[width=.95\linewidth]{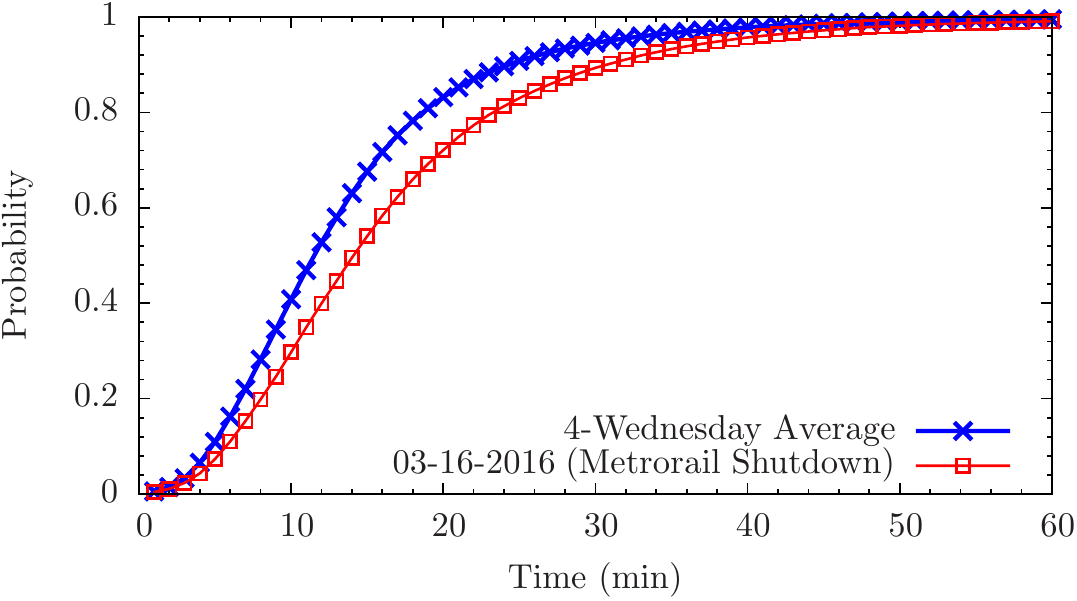} \caption{Changes in Trip Time Statistics.}
\label{fig:dc_taxidata_1Y_grp_TripTime_Metro16_Comp_Sub_PDF_CDF} 
\end{subfigure}

\caption{Changes of Taxi Trips due to Metrorail Shutdown on 03-16-2016.}
\label{fig:dc_taxidata_1Y_metro_1603_shutdown}
\end{figure}

Here we investigate the role of taxi system as one mode of multimodal transportation on preserving resilience of a city-scale transportation network.
Fig. \ref{fig:dc_taxidata_1Y_metro_1603_shutdown} shows the impact of this major disruption on taxi trips by comparing the results on 03-16-2016 and the results of a four-Wednesday average (two Wednesdays before and two Wednesdays after the disruption). The volume of taxi trips on 03-16-2016 increased 23.2\% in comparison with that of the four-Wednesday average (see Fig.~\ref{fig:dc_taxidata_Metro16RangeVsDate_hod_Week}), while the median time of taxi trips increased 22.4\% from 11.53 to 14.11 minutes (see Fig.~\ref{fig:dc_taxidata_1Y_grp_TripTime_Metro16_Comp_Sub_PDF_CDF}). 
These elastic changes reflect that the taxi system played a significant function on help preserving the transportation resilience as such a major disruption occurred. 

In January of 2016, the blizzard ``Snowzilla'' \citep{US2016Blizzard} caused significant snowfall and blizzard conditions, and a snow emergency was declared in the DC area. The Washington Metropolitan Area Transit Authority (WMATA) suspended all Metrorail and Metrobus service between 01-23-2016 and 01-24-2016. Capital Bikeshare closed its stations between 01-23-2016 and 01-26-2016. 
Fig. \ref{fig:dc_taxidata_1Y_grp_Hour_2016-01_Snowstorm_Comp_d} shows the taxi trip volumes in the blizzard week (from 01-21-2016 to 01-28-2016, in red color) and a normal week (from 01-07-2016 to 01-14-2016, in blue color). For indicating temporal resilience, Fig.~\ref{fig:dc_taxidata_1Y_grp_Hour_2016-01_Snowstorm_Comp_ratio} shows the hourly ratios $r_{h}$ between the two weeks as defined in Eq.~\ref{eq:ratio_taxi_trip_volumes},
\begin{equation}
r_{h} = \frac{\text{Taxi Volume for Each Hour in the Blizzard Week}}{\text{Taxi Volume for Each Hour in the Normal Week}}.
\label{eq:ratio_taxi_trip_volumes}
\end{equation}
Figs. \ref{fig:dc_taxidata_1Y_geo_2016-01-21_pick_r3}--\ref{fig:dc_taxidata_1Y_geo_2016-01-28_pick_r3} give the spatial distributions of taxi pickups for each day in the blizzard week, where pickup locations are decomposed with $k$-digit grids using $k=3$ (see Definition \ref{def:rDGD}). In Figs.~\ref{fig:dc_taxidata_1Y_geo_2016-01-21_pick_r3}--\ref{fig:dc_taxidata_1Y_geo_2016-01-28_pick_r3}, each circle represents the center of a pickup location decomposed with $3$-digit grid, and its color represents the range of daily taxi pickup rate in $\operatorname{log}_{10}$ corresponding to the pickup location.

\afterpage{


\begin{figure} [H]
\centering
\begin{subfigure}{.95\textwidth} \centering \includegraphics[width=.95\linewidth]{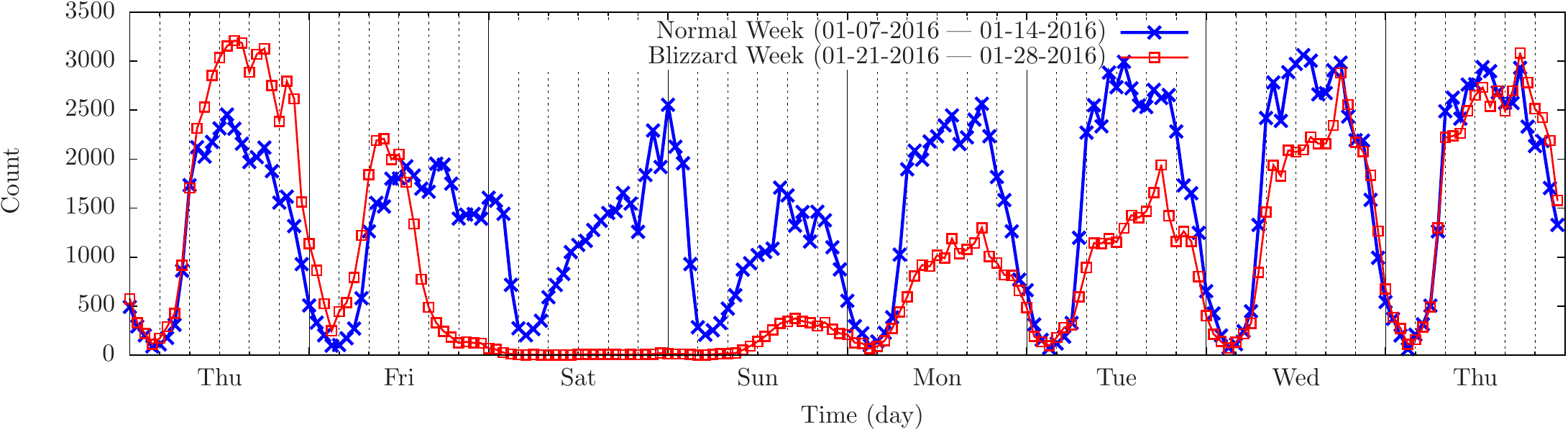} \caption{Taxi Pickup Volumes in the Blizzard Week (01-21 to 01-28) and a Normal Week (01-07 to 01-14).}
\label{fig:dc_taxidata_1Y_grp_Hour_2016-01_Snowstorm_Comp_d} 
\end{subfigure}

\begin{subfigure}{.95\textwidth} \centering \includegraphics[width=.95\linewidth]{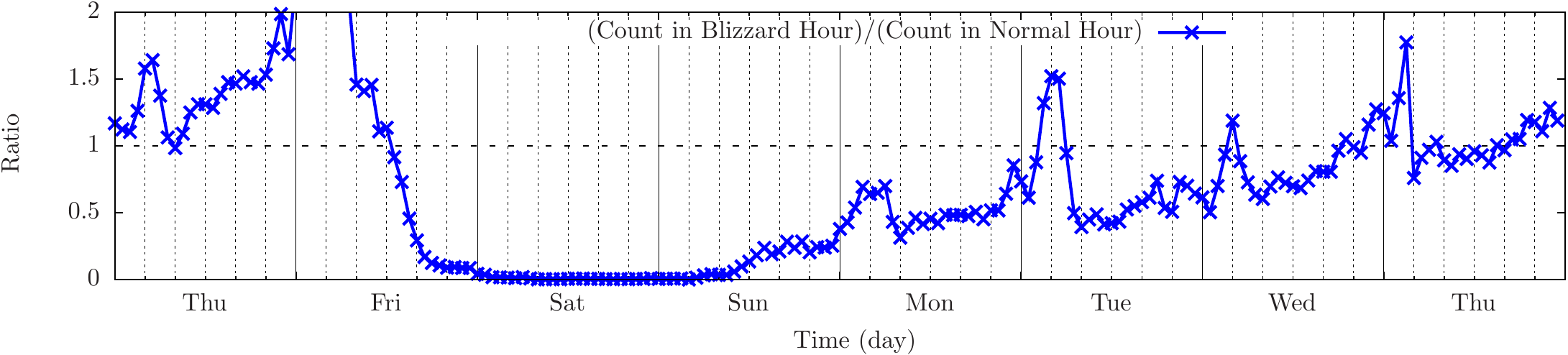} \caption{Hourly Ratios of Taxi Pickup Volumes between the Blizzard Week and the Normal Week.}
\label{fig:dc_taxidata_1Y_grp_Hour_2016-01_Snowstorm_Comp_ratio} 
\end{subfigure}
\caption{Temporal Taxi Pickup Distributions: Comparison between the Blizzard Week and a Normal Week in 2016.}
\label{fig:dc_taxidata_1Y_grp_Hour_2016-01_Snowstorm_Comp} 
\end{figure}

\begin{figure} [H]
\centering
\begin{subfigure}{.244\textwidth} \centering \includegraphics[width=.99\linewidth]{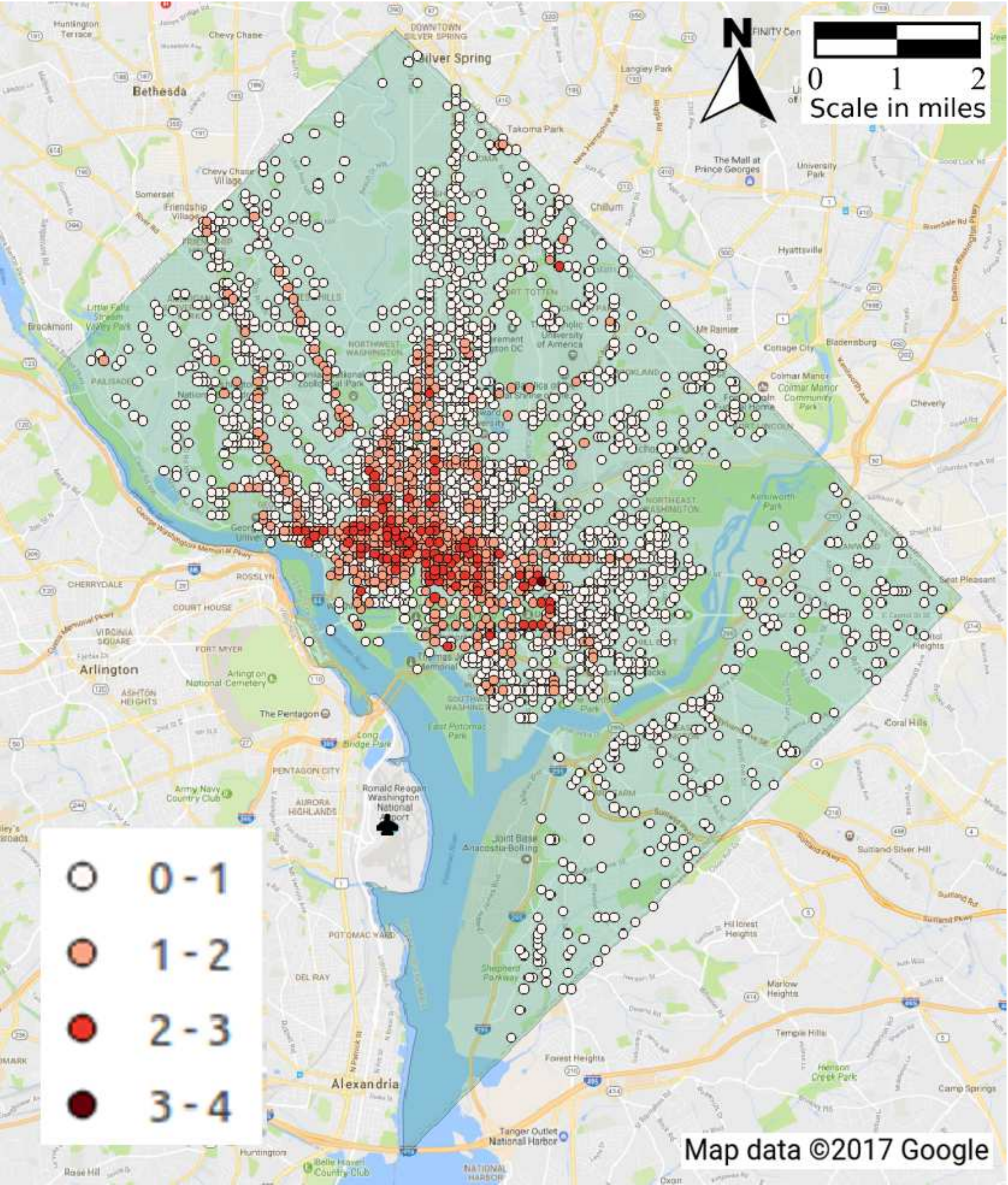} \caption{01-21-2016: $r_{g3}$=1.07}
\label{fig:dc_taxidata_1Y_geo_2016-01-21_pick_r3} 
\end{subfigure}
\begin{subfigure}{.244\textwidth} \centering \includegraphics[width=.99\linewidth]{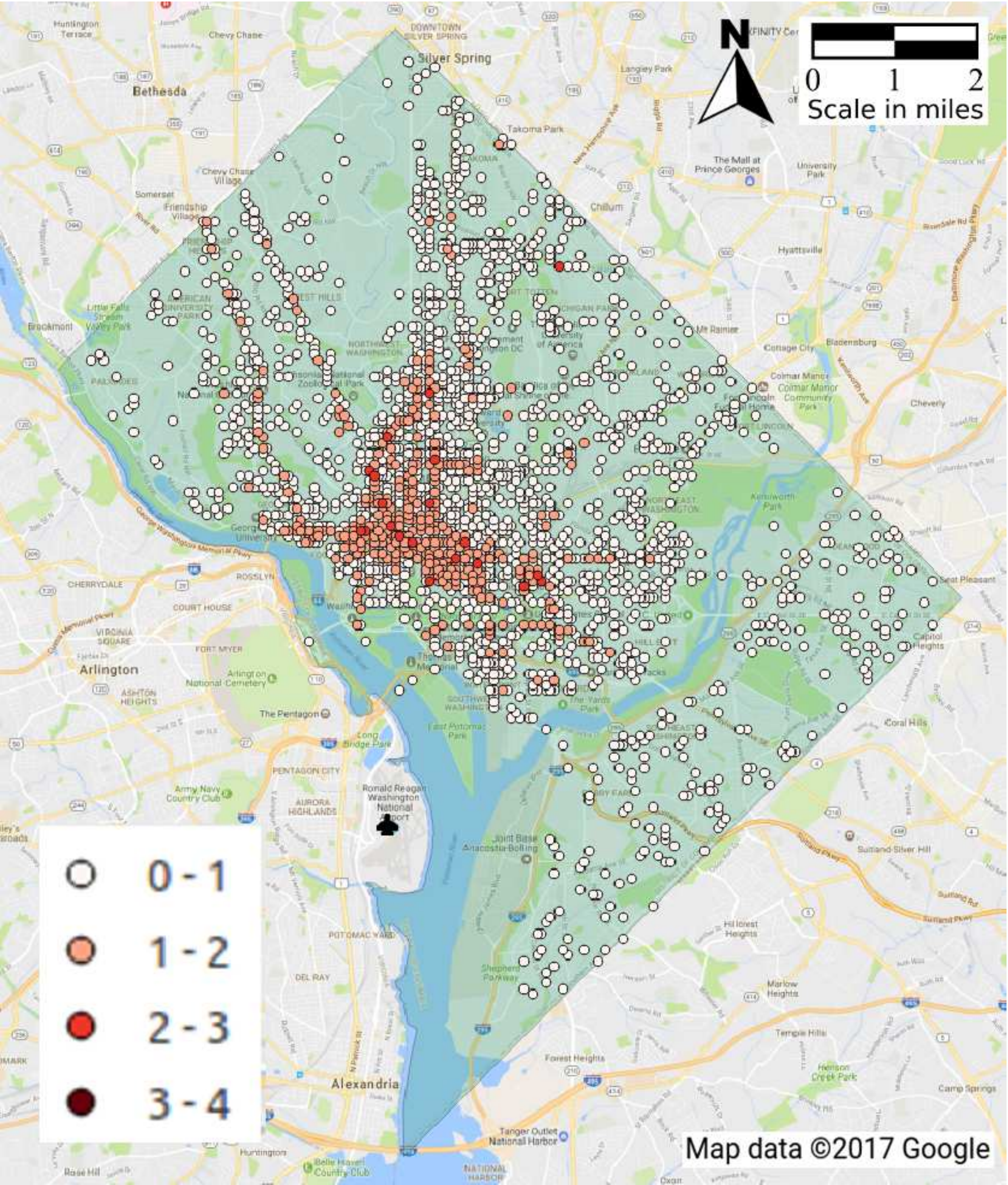} \caption{01-22-2016: $r_{g3}$=0.94}
\label{fig:dc_taxidata_1Y_geo_2016-01-22_pick_r3} 
\end{subfigure}
\begin{subfigure}{.244\textwidth} \centering \includegraphics[width=.99\linewidth]{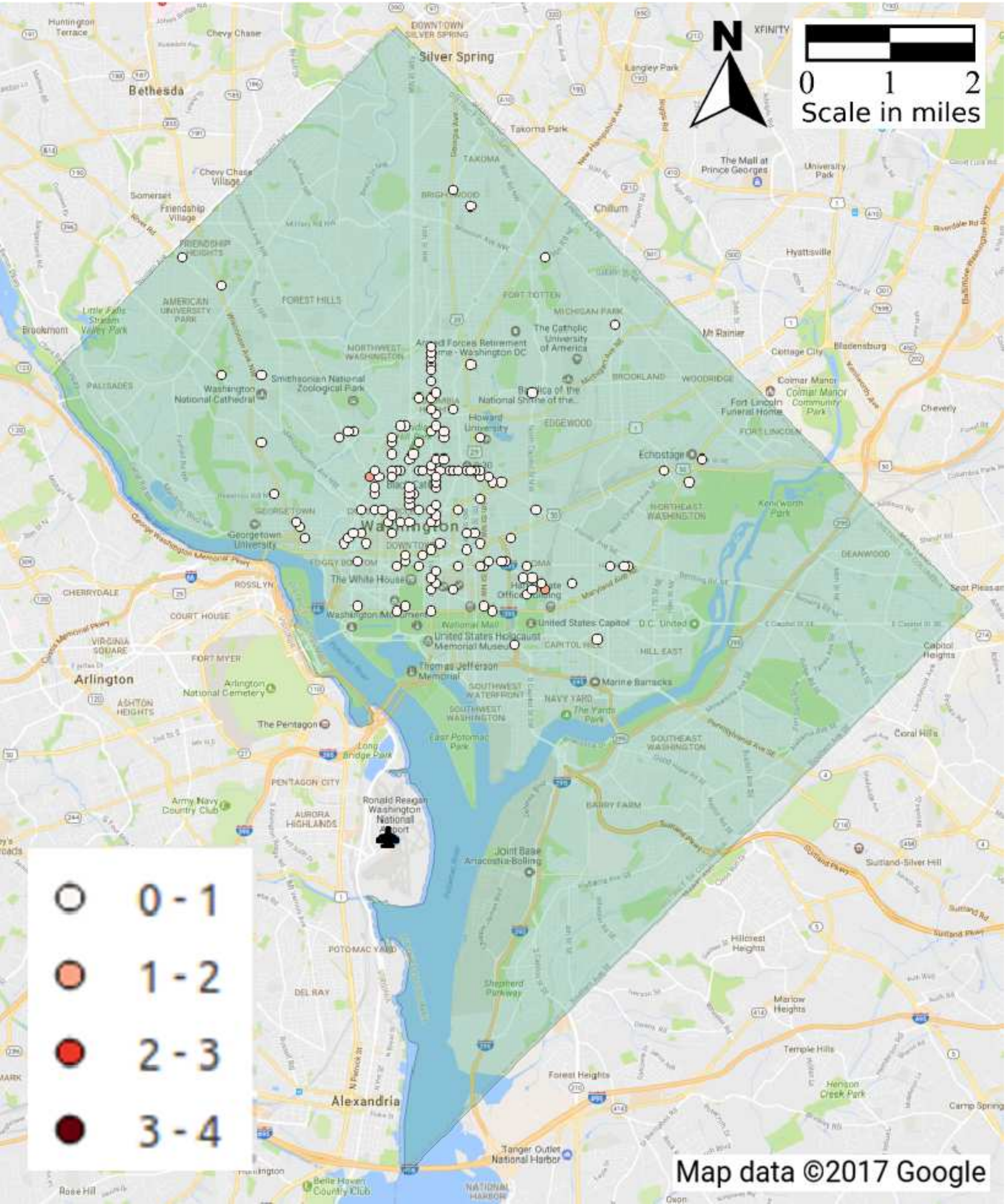} \caption{01-23-2016: $r_{g3}$=0.07}
\label{fig:dc_taxidata_1Y_geo_2016-01-23_pick_r3} 
\end{subfigure}
\begin{subfigure}{.244\textwidth} \centering \includegraphics[width=.99\linewidth]{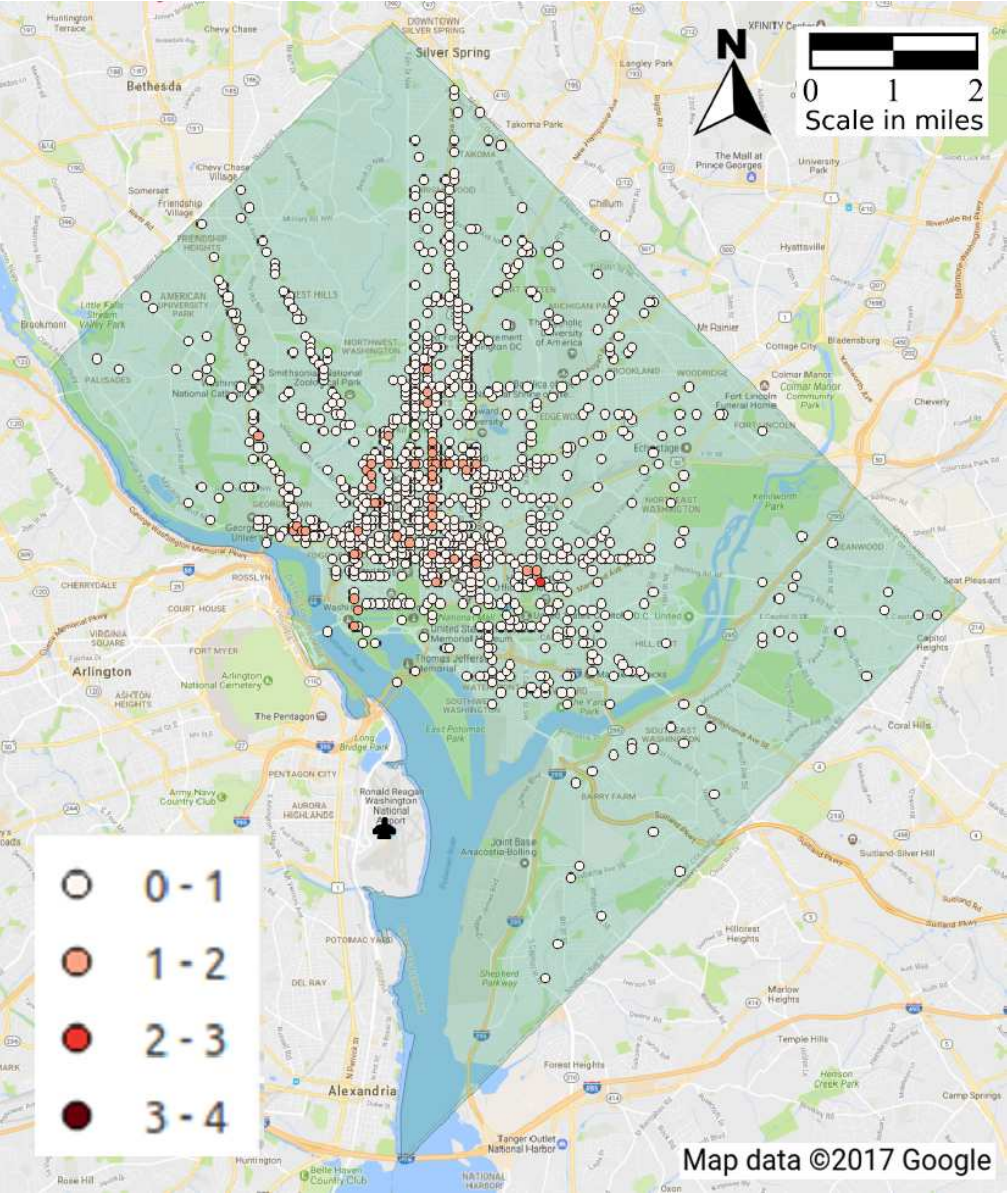} \caption{01-24-2016: $r_{g3}$=0.44}
\label{fig:dc_taxidata_1Y_geo_2016-01-24_pick_r3} 
\end{subfigure}

\begin{subfigure}{.244\textwidth} \centering \includegraphics[width=.99\linewidth]{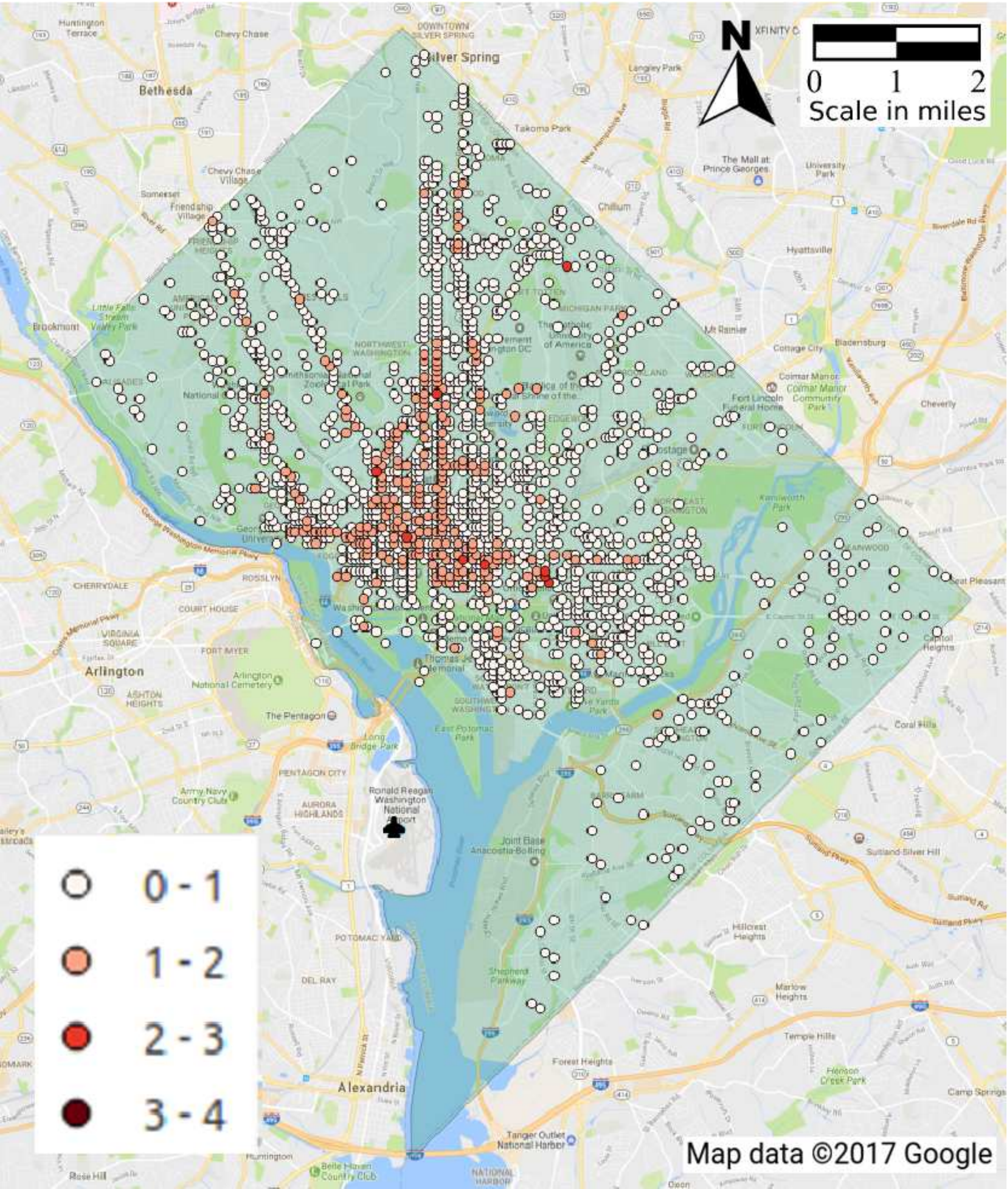} \caption{01-25-2016: $r_{g3}$=0.78}
\label{fig:dc_taxidata_1Y_geo_2016-01-25_pick_r3} 
\end{subfigure}
\begin{subfigure}{.244\textwidth} \centering \includegraphics[width=.99\linewidth]{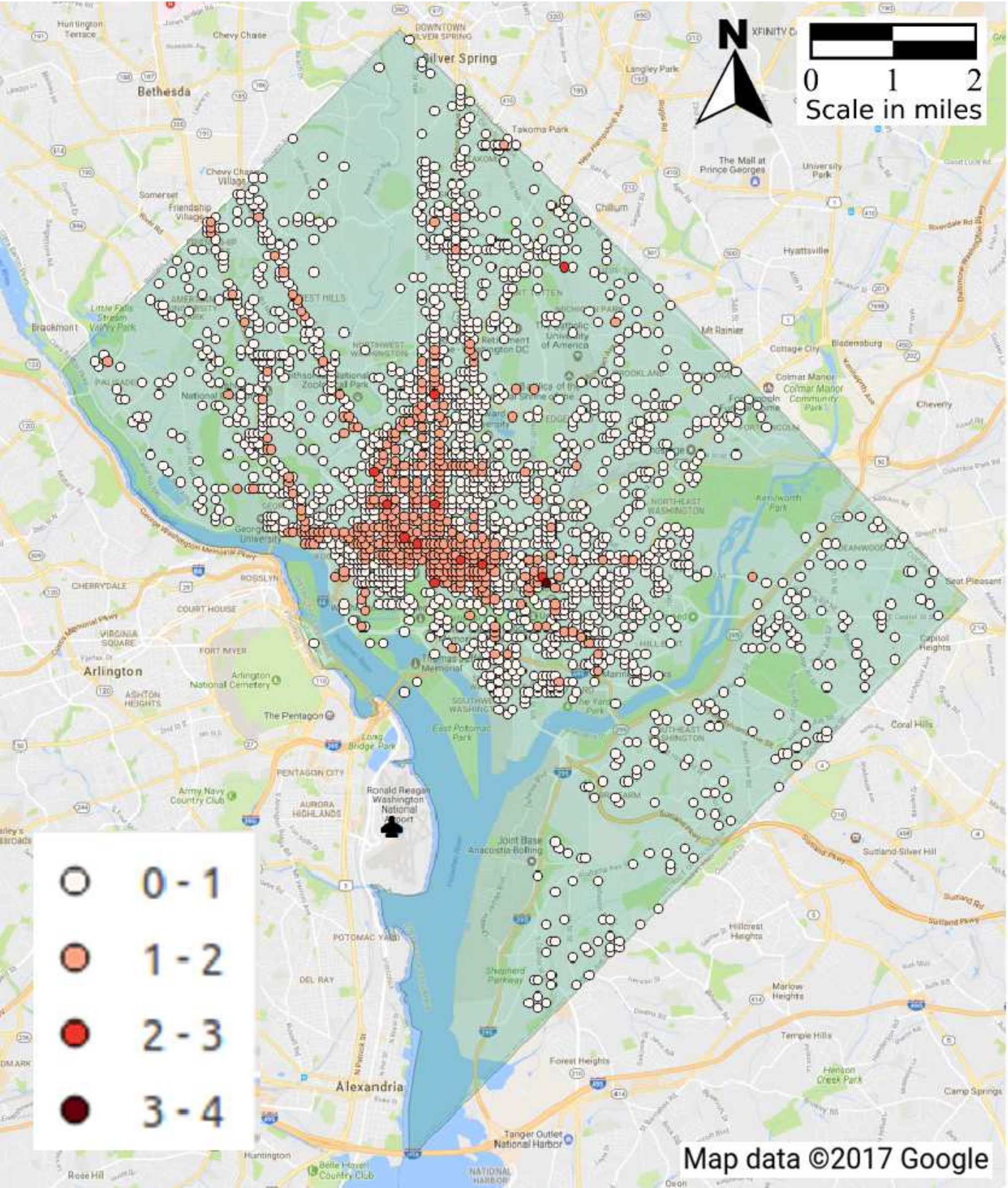} \caption{01-26-2016: $r_{g3}$=0.90}
\label{fig:dc_taxidata_1Y_geo_2016-01-26_pick_r3} 
\end{subfigure}
\begin{subfigure}{.244\textwidth} \centering \includegraphics[width=.99\linewidth]{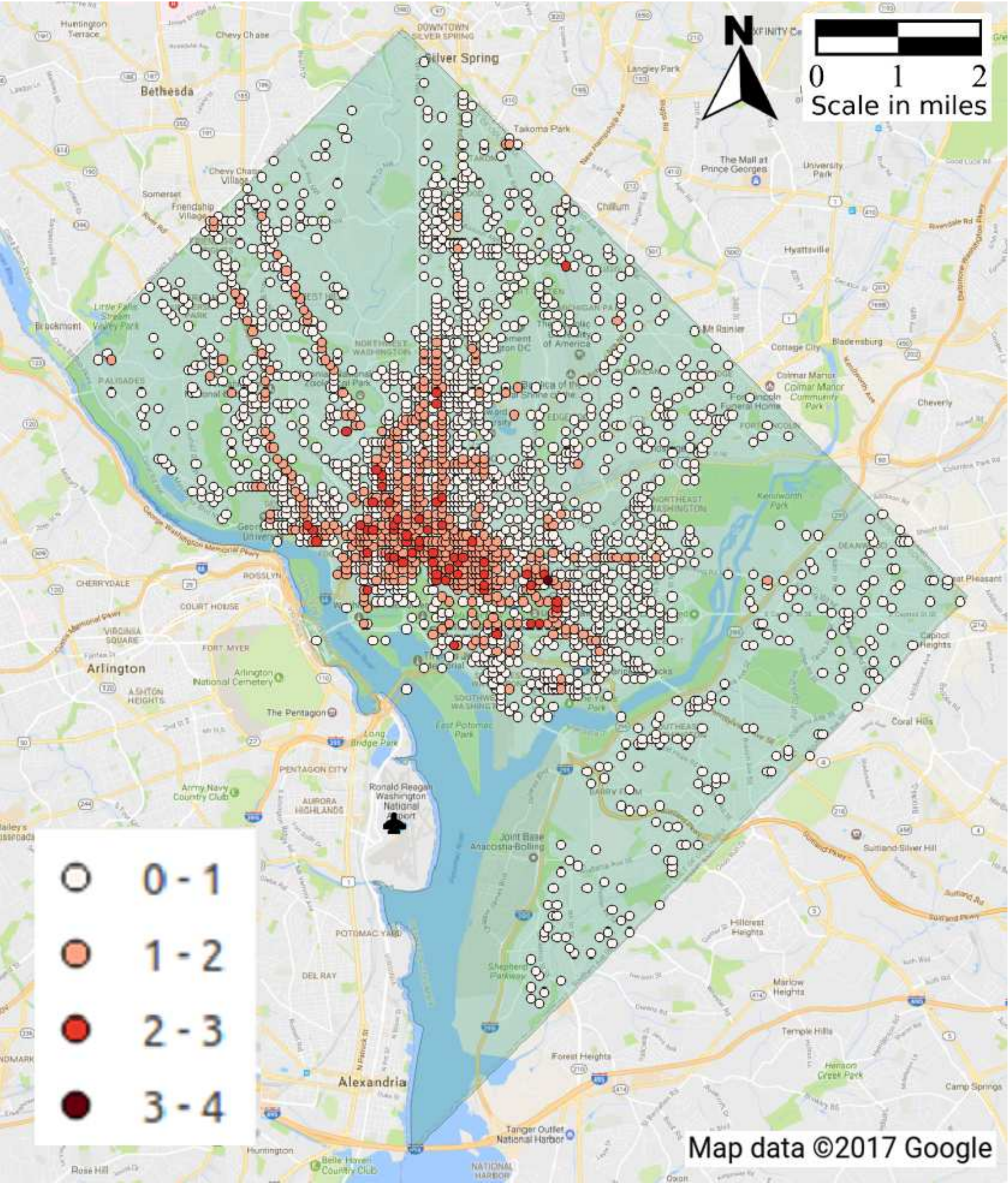} \caption{01-27-2016: $r_{g3}$=1.02}
\label{fig:dc_taxidata_1Y_geo_2016-01-27_pick_r3} 
\end{subfigure}
\begin{subfigure}{.244\textwidth} \centering \includegraphics[width=.99\linewidth]{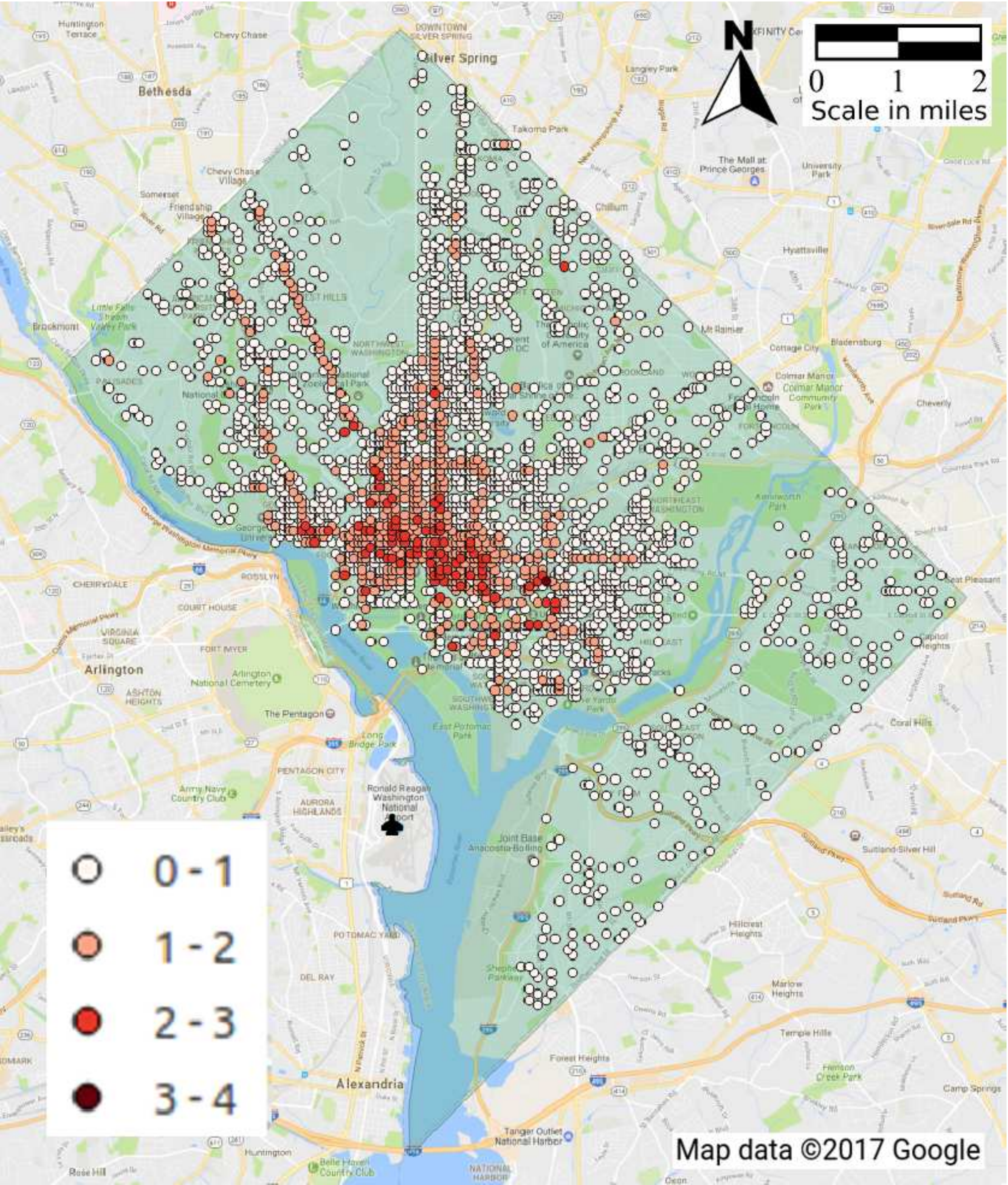} \caption{01-28-2016: $r_{g3}$=1.07}
\label{fig:dc_taxidata_1Y_geo_2016-01-28_pick_r3} 
\end{subfigure}

\caption{Spatial Taxi Pickup Distributions and $r_{g3}$ Values in the Week of January 2016 Blizzard {\em Snowzilla}.}
\label{fig:dc_taxidata_1Y_metro_1601_shutdown}
\end{figure}
}

As shown in Fig. \ref{fig:dc_taxidata_1Y_grp_Hour_2016-01_Snowstorm_Comp_d}, the taxi ridership suddenly dropped down from the afternoon of Friday (01-22-2016). 
On Saturday (01-23-2016), though only a few taxi trips were taken (see Fig.  \ref{fig:dc_taxidata_1Y_grp_Hour_2016-01_Snowstorm_Comp_d}), the service of taxi transportation system covered a substantial region of DC (see Fig. \ref{fig:dc_taxidata_1Y_geo_2016-01-23_pick_r3}). On Sunday (01-24-2016), there were a large amount (13.7\% of the normal state) of taxi trips (see Fig.  \ref{fig:dc_taxidata_1Y_grp_Hour_2016-01_Snowstorm_Comp_d}), and the taxi system recovered its function in the major urban areas of DC (see Fig. \ref{fig:dc_taxidata_1Y_geo_2016-01-24_pick_r3})). During these two weekend days (01-23-2016 and 01-24-2016), the taxi system worked as a single transportation mode, because Metrorail, Metrobus, and Capital Bikeshare did not operate. The volume of taxi trips returned to 47.5\% of its normal state on Monday (01-25-2016), and it gradually reached to the complete normal state on Thursday (01-28-2016), see Fig. \ref{fig:dc_taxidata_1Y_grp_Hour_2016-01_Snowstorm_Comp_d}. During the recovery period, the increase in the traffic demands using the taxi system was partly contributed by the major customer sources of two other modes in the multimodal transportation system, mainly the people commuting by rail and by air~\citep{schaller2005regression}. 
Concerning the scenario, traditional measurement indices on resilience, such as peak disruption and recovery time~\citep{donovan2017empirically}, can be acquired from Fig. \ref{fig:dc_taxidata_1Y_grp_Hour_2016-01_Snowstorm_Comp_ratio} as the ratio drops below 1. The two indices however may not be sufficient to describe the full spatio-temporal aspects of transportation resilience in a multimodal environment. Therefore, in Fig \ref{fig:dc_taxidata_1Y_metro_1601_shutdown}, we not only show the spatial distributions of taxi pickups for each day in the blizzard week, but also compare the taxi spatial accessibility for each day between the blizzard and the normal weeks through computing $r_{g3}$ as defined in Eq.~\ref{eq:ratio_taxi_accessibility} for indicating spatial resilience,
\begin{equation}
r_{g3} = \frac{\text{Numbers of 3-Digit Grids Accessed for Each day in the Blizzard Week}}{\text{Numbers of 3-Digit Grids Accessed for Each day in the Normal Week}},
\label{eq:ratio_taxi_accessibility}
\end{equation}
which represents the ratio of the total numbers of 3-digit grids where taxi accessed between the blizzard and the normal weeks. The $r_{g3}$ values are respectively 0.07 and 0.44 on 01-23-2016 and 01-24-2016. Notice that we did not count any walking distance in the computation, and the ratios would be higher if walking distances were considered inside. The value of $r_{g3}$ at peak disruption, 0.07, represents the spatial accessibility of surface transportation modes in the city under the worst condition in such an extreme event. For the taxi system, the recovery time is 3 days if using a threshold $r_{g3}\geq 0.9$, regarding the resilience in spatial accessibility. 


\section{Trip Patterns and Land Use} \label{sec:landuse}

\begin{figure} [p]
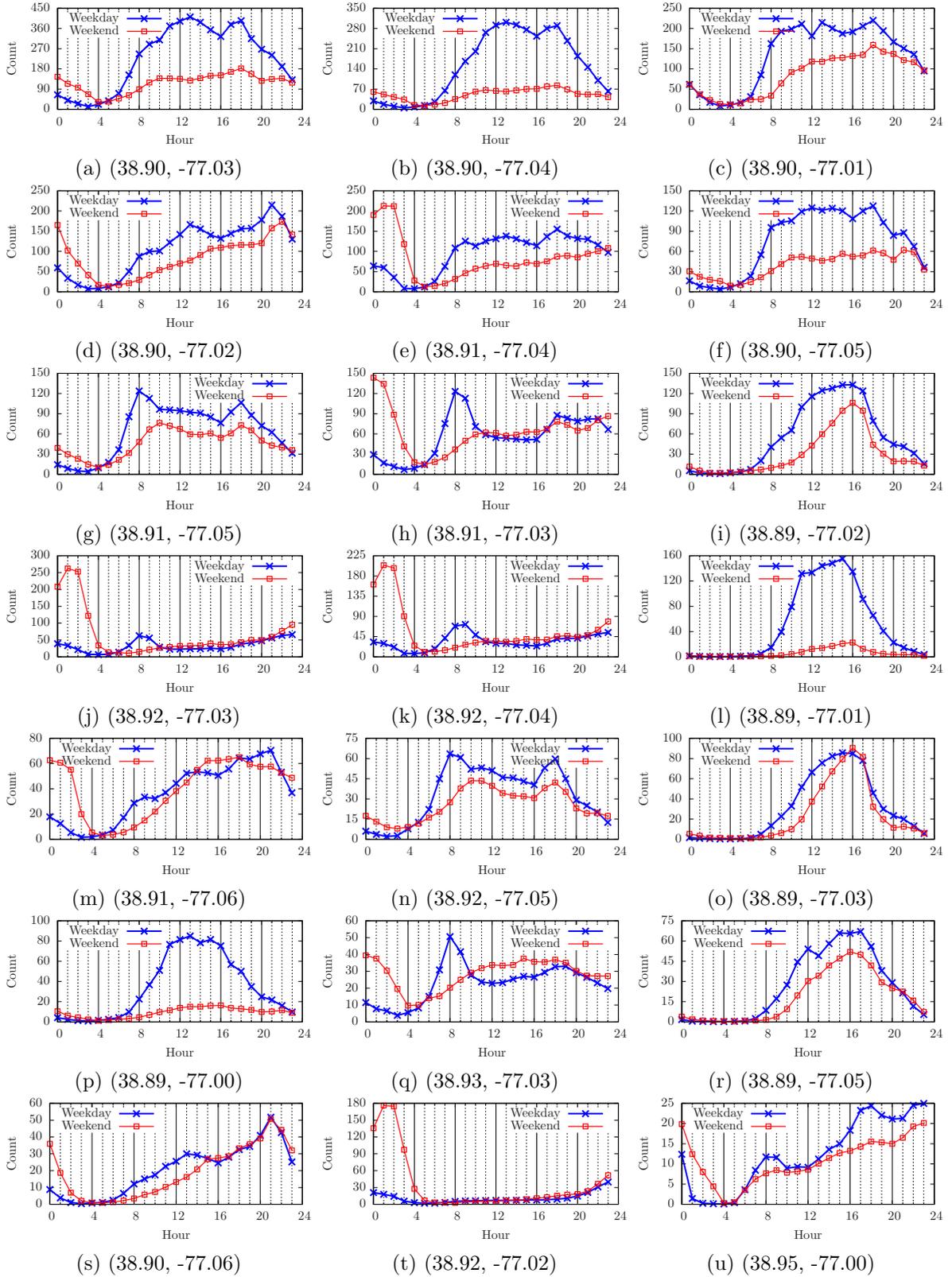

\centering

\begin{subfigure}{.310\textwidth} \centering \includegraphics[width=.99\linewidth]{img2/{{dc_taxidata_1Y_r2_38.90,-77.03_pick_hod_Week}}} \caption{(38.90, -77.03)} \label{fig:dc_taxidata_1Y_r2_38.90,-77.03_pick_hod_Week} \end{subfigure}
\begin{subfigure}{.310\textwidth} \centering \includegraphics[width=.99\linewidth]{img2/{{dc_taxidata_1Y_r2_38.90,-77.04_pick_hod_Week}}} \caption{(38.90, -77.04)} \label{fig:dc_taxidata_1Y_r2_38.90,-77.04_pick_hod_Week} \end{subfigure}
\begin{subfigure}{.310\textwidth} \centering \includegraphics[width=.99\linewidth]{img2/{{dc_taxidata_1Y_r2_38.90,-77.01_pick_hod_Week}}} \caption{(38.90, -77.01)} \label{fig:dc_taxidata_1Y_r2_38.90,-77.01_pick_hod_Week} \end{subfigure}

\begin{subfigure}{.310\textwidth} \centering \includegraphics[width=.99\linewidth]{img2/{{dc_taxidata_1Y_r2_38.90,-77.02_pick_hod_Week}}} \caption{(38.90, -77.02)} \label{fig:dc_taxidata_1Y_r2_38.90,-77.02_pick_hod_Week} \end{subfigure}
\begin{subfigure}{.310\textwidth} \centering \includegraphics[width=.99\linewidth]{img2/{{dc_taxidata_1Y_r2_38.91,-77.04_pick_hod_Week}}} \caption{(38.91, -77.04)} \label{fig:dc_taxidata_1Y_r2_38.91,-77.04_pick_hod_Week} \end{subfigure}
\begin{subfigure}{.310\textwidth} \centering \includegraphics[width=.99\linewidth]{img2/{{dc_taxidata_1Y_r2_38.90,-77.05_pick_hod_Week}}} \caption{(38.90, -77.05)} \label{fig:dc_taxidata_1Y_r2_38.90,-77.05_pick_hod_Week} \end{subfigure}

\begin{subfigure}{.310\textwidth} \centering \includegraphics[width=.99\linewidth]{img2/{{dc_taxidata_1Y_r2_38.91,-77.05_pick_hod_Week}}} \caption{(38.91, -77.05)} \label{fig:dc_taxidata_1Y_r2_38.91,-77.05_pick_hod_Week} \end{subfigure}
\begin{subfigure}{.310\textwidth} \centering \includegraphics[width=.99\linewidth]{img2/{{dc_taxidata_1Y_r2_38.91,-77.03_pick_hod_Week}}} \caption{(38.91, -77.03)} \label{fig:dc_taxidata_1Y_r2_38.91,-77.03_pick_hod_Week} \end{subfigure}
\begin{subfigure}{.310\textwidth} \centering \includegraphics[width=.99\linewidth]{img2/{{dc_taxidata_1Y_r2_38.89,-77.02_pick_hod_Week}}} \caption{(38.89, -77.02)} \label{fig:dc_taxidata_1Y_r2_38.89,-77.02_pick_hod_Week} \end{subfigure}

\begin{subfigure}{.310\textwidth} \centering \includegraphics[width=.99\linewidth]{img2/{{dc_taxidata_1Y_r2_38.92,-77.03_pick_hod_Week}}} \caption{(38.92, -77.03)} \label{fig:dc_taxidata_1Y_r2_38.92,-77.03_pick_hod_Week} \end{subfigure}
\begin{subfigure}{.310\textwidth} \centering \includegraphics[width=.99\linewidth]{img2/{{dc_taxidata_1Y_r2_38.92,-77.04_pick_hod_Week}}} \caption{(38.92, -77.04)} \label{fig:dc_taxidata_1Y_r2_38.92,-77.04_pick_hod_Week} \end{subfigure}
\begin{subfigure}{.310\textwidth} \centering \includegraphics[width=.99\linewidth]{img2/{{dc_taxidata_1Y_r2_38.89,-77.01_pick_hod_Week}}} \caption{(38.89, -77.01)} \label{fig:dc_taxidata_1Y_r2_38.89,-77.01_pick_hod_Week} \end{subfigure}

\begin{subfigure}{.310\textwidth} \centering \includegraphics[width=.99\linewidth]{img2/{{dc_taxidata_1Y_r2_38.91,-77.06_pick_hod_Week}}} \caption{(38.91, -77.06)} \label{fig:dc_taxidata_1Y_r2_38.91,-77.06_pick_hod_Week} \end{subfigure}
\begin{subfigure}{.310\textwidth} \centering \includegraphics[width=.99\linewidth]{img2/{{dc_taxidata_1Y_r2_38.92,-77.05_pick_hod_Week}}} \caption{(38.92, -77.05)} \label{fig:dc_taxidata_1Y_r2_38.92,-77.05_pick_hod_Week} \end{subfigure}
\begin{subfigure}{.310\textwidth} \centering \includegraphics[width=.99\linewidth]{img2/{{dc_taxidata_1Y_r2_38.89,-77.03_pick_hod_Week}}} \caption{(38.89, -77.03)} \label{fig:dc_taxidata_1Y_r2_38.89,-77.03_pick_hod_Week} \end{subfigure}

\begin{subfigure}{.310\textwidth} \centering \includegraphics[width=.99\linewidth]{img2/{{dc_taxidata_1Y_r2_38.89,-77.00_pick_hod_Week}}} \caption{(38.89, -77.00)} \label{fig:dc_taxidata_1Y_r2_38.89,-77.00_pick_hod_Week} \end{subfigure}
\begin{subfigure}{.310\textwidth} \centering \includegraphics[width=.99\linewidth]{img2/{{dc_taxidata_1Y_r2_38.93,-77.03_pick_hod_Week}}} \caption{(38.93, -77.03)} \label{fig:dc_taxidata_1Y_r2_38.93,-77.03_pick_hod_Week} \end{subfigure}
\begin{subfigure}{.310\textwidth} \centering \includegraphics[width=.99\linewidth]{img2/{{dc_taxidata_1Y_r2_38.89,-77.05_pick_hod_Week}}} \caption{(38.89, -77.05)} \label{fig:dc_taxidata_1Y_r2_38.89,-77.05_pick_hod_Week} \end{subfigure}

\begin{subfigure}{.310\textwidth} \centering \includegraphics[width=.99\linewidth]{img2/{{dc_taxidata_1Y_r2_38.90,-77.06_pick_hod_Week}}} \caption{(38.90, -77.06)} \label{fig:dc_taxidata_1Y_r2_38.90,-77.06_pick_hod_Week} \end{subfigure}
\begin{subfigure}{.310\textwidth} \centering \includegraphics[width=.99\linewidth]{img2/{{dc_taxidata_1Y_r2_38.92,-77.02_pick_hod_Week}}} \caption{(38.92, -77.02)} \label{fig:dc_taxidata_1Y_r2_38.92,-77.02_pick_hod_Week} \end{subfigure}
\begin{subfigure}{.310\textwidth} \centering \includegraphics[width=.99\linewidth]{img2/{{dc_taxidata_1Y_r2_38.95,-77.00_pick_hod_Week}}} \caption{(38.95, -77.00)} \label{fig:dc_taxidata_1Y_r2_38.95,-77.00_pick_hod_Week} \end{subfigure}

\caption{Temporal Patterns of the Top 21 highest 2-Digit Grids Ranking in Taxi Pickup Volumes.}
\label{fig:dc_taxidata_1Y_r2_pick_hod_Week}
\end{figure}

Taxi trips contain massive spatial and temporal information on human mobility \citep{castro2013taxi,tang2015uncovering}. By fusing multiple sources of participatory sensing data \citep{Xie2015}, we can better understand social dynamics and the land use in different regions of a city, which not only provides an essential support to the management and maintenance on the public infrastructure and resources of a city, but also enables transportation planning in a timely fashion.
Fig. \ref{fig:dc_taxidata_1Y_r2_pick_hod_Week} shows the comparisons of temporal patterns between weekday and weekend for the top 21 highest 2-digit grids in the ranking of the volumes of taxi pickups. As shown in Fig. \ref{fig:dc_taxidata_1Y_r2_pick_hod_Week}, temporal patterns of human activity are divergent among different grids.    

A pattern of human activity for a region expresses an aggregation effect integrating all the activities taken by all the individual persons at all the places in the region. We use the Foursquare API to obtain a large sampling on population. Foursquare is a location-based social network (LBSN). In Foursquare, each place is called a {\em venue}, where users can check in their activities. The venues are organized into hierarchical categories based on their supported activities and functions. There are total 10 categories at the root level, which are respectively 1) Arts \& Entertainment, 2) College \& University, 3) Event, 4) Food, 5) Nightlife Spot, 6) Outdoors \& Recreation, 7) Professional \& Other Places, 8) Residence, 9) Shop \& Service, and 10) Travel \& Transport. 

The numbers of check-ins are collected every half hour for all the venues within the whole studied area of the Foursquare dataset. Using the data, for each venue $v$, we can calculate its average check-in rate $r_{c(v)}$ as
\begin{equation}
r_{c(v)} = (s_{c(v)}^{max}-s_{c(v)}^{min})/(t_{c(v)}^{max}-t_{c(v)}^{min}),
\end{equation}
where $t_{c(v)}^{min}$ and $t_{c(v)}^{max}$ are the check-in time at beginning and at last, $s_{c(v)}^{max}$ and $s_{c(v)}^{min}$ are the check-in counts at beginning and at last. We discard the venues with  $t_{c(v)}^{max}=t_{c(v)}^{min}$, as these venues have very low check-in rates (since no any change in the check-in counts during the long collection period).

Let $q_{c(v)}$ be the root category of each venue $v$, we can compute the weight of each root category $Q$ for each region $R$ by summing up $q_{c(v)}$ over all the venues in the region $R$ as
\begin{equation}
w_{(R, Q)}=\sum_{v \in R, q_{c(v)}=Q} r_{c(v)},
\end{equation}
and the normalized weight $\hat{w}_{(R, Q)}$ is
\begin{equation}
\hat{w}_{(R, Q)}=w_{(R, Q)}/\sum_{\forall Q} w_{(R, Q)}. 
\end{equation}
Table \ref{tab:venue_weights_regions} lists the normalized weights in percentage for all root categories of venues in different regions, where Regions A, B, C, D, E, F are respectively corresponding to the grids depicted in Figs.
\ref{fig:dc_taxidata_1Y_r2_38.90,-77.03_pick_hod_Week}, 
\ref{fig:dc_taxidata_1Y_r2_38.91,-77.04_pick_hod_Week}, 
\ref{fig:dc_taxidata_1Y_r2_38.90,-77.01_pick_hod_Week}, 
\ref{fig:dc_taxidata_1Y_r2_38.92,-77.03_pick_hod_Week}, 
\ref{fig:dc_taxidata_1Y_r2_38.89,-77.02_pick_hod_Week}, 
\ref{fig:dc_taxidata_1Y_r2_38.89,-77.01_pick_hod_Week}. These regions are considered since they have rather high weights in some root categories. Compared to traditional zoning and land use maps, the participatory sensing data can provide detailed proportions of lane use categories in different regions. 

\begin{table*} [h]
\centering  \caption{Normalized Weights (in \%) for All Root Categories of Venues in Different Regions.}
  \label{tab:venue_weights_regions}
\begin{tabular}{|l|l|c|c|c|c|c|c|c|} \hline 

No. & Venue Root Category & A & B & C & D & E & F  \tabularnewline \hline 
1 & Arts \& Entertainment & 2.61 & 1.82 & 1.24 & 3.60 & 37.37 & 6.88 \\ \hline 
2 & College \& University & 0.93 & 0.98 & 1.89 & 0.09 & 0.13 & 0.00 \\ \hline 
3 & Event & 0.44 & 0.00 & 0.01 & 0.00 & 0.00 & 0.00 \\ \hline 
4 & Food & 26.6 & 33.53 & 17.66 & 35.42 & 16.08 & 7.44 \\ \hline 
5 & Nightlife Spot & 14.01 & 14.24 & 0.63 & 18.97 & 0.91 & 0.66 \\ \hline 
6 & Outdoors \& Recreation & 12.72 & 7.85 & 2.41 & 9.27 & 18.48 & 10.58 \\ \hline 
7 & Professional \& Other Places & 21.26 & 19.30 & 13.72 & 5.85 & 13.71 & 69.98 \\ \hline 
8 & Residence & 1.17 & 2.04 & 2.40 & 4.04 & 0.20 & 0.00 \\ \hline 
9 & Shop \& Service & 8.14 & 11.61 & 6.19 & 18.23 & 1.64 & 1.65 \\ \hline 
10 & Travel \& Transport & 12.13 & 8.64 & 53.84 & 4.54 & 11.47 & 2.82 \\ \hline 

\end{tabular}  
\end{table*}

From Fig. \ref{fig:dc_taxidata_1Y_r2_pick_hod_Week}, we can identify four basic types of regional taxi trip demand. Here we explain each type of taxi demand, and find the corresponding information in Table \ref{tab:venue_weights_regions} on land use. The first type is Travel \& Transportation hubs, including major train stations and airports \citep{ferreira2013visual}. Its example is Regions C, which has 53.84\% venues in the Category No. 10 as shown in Table \ref{tab:venue_weights_regions}. The region contains the Union Station, a major transportation hub of DC. As shown in Fig. \ref{fig:dc_taxidata_1Y_r2_38.90,-77.01_pick_hod_Week}, the pickup volume in Region C is high during the daytime and early night time. The second type is Professional Places, such as Region F, which have 69.98\% venues in Category No. 7, as shown in Table \ref{tab:venue_weights_regions}. This region contains the Capital Hill. The pickup flow patterns include high volumes during the daytime in weekday, and very low volumes in weekend, as shown in Fig. \ref{fig:dc_taxidata_1Y_r2_38.89,-77.02_pick_hod_Week}. A similar region is shown in Fig. \ref{fig:dc_taxidata_1Y_r2_38.89,-77.00_pick_hod_Week}. The third type is Recreation \& Entertainment, such as Region E, which have 37.37\% and 18.48\% in Categories No. 1 and No. 6, as shown in Table \ref{tab:venue_weights_regions}. The pickup flow patterns include a main peak at afternoon, both in weekday and weekend, as shown in Fig. \ref{fig:dc_taxidata_1Y_r2_38.89,-77.02_pick_hod_Week}. Similar patterns can be found in Figs. \ref{fig:dc_taxidata_1Y_r2_38.89,-77.03_pick_hod_Week} and  \ref{fig:dc_taxidata_1Y_r2_38.89,-77.05_pick_hod_Week}. The fourth type is Nightlife Spot, which has more venues in Categories No. 4, 5 and 9. An example is Region D, which is located near the U Street Corridor. Its main feature in pickup flow pattern is a high volume peak during 0-4 AM of weekend, as shown in Fig. \ref{fig:dc_taxidata_1Y_r2_38.92,-77.03_pick_hod_Week}. Similar patterns can be identified in Figs. \ref{fig:dc_taxidata_1Y_r2_38.92,-77.04_pick_hod_Week} and \ref{fig:dc_taxidata_1Y_r2_38.92,-77.02_pick_hod_Week}. 

Trip demand for some regions might be a mix of multiple basis flows~\citep{peng2012collective}, for example, Regions A and B. Region A has the highest total number of taxi pickups among all 2-digit grids, and it has a mixed form of venues including Food, Professional Places but without a single dominant Category. Region B is near Dupont Circle, which is a result of mixing more forms of venues in Categories 4, 9 and 5, and the mixed effect of multiple venue categories is shown to have attracted more people for night life in weekends, see Fig. \ref{fig:dc_taxidata_1Y_r2_38.91,-77.04_pick_hod_Week}. 

In summary, the results indicate that the significant relations between trip patterns and land use types can be uncovered by fusing taxi trip data and participatory sensing data. The findings are able to support a better comprehension on the multi-facet interactions between human mobility dynamics and urban spatial structures.

\section{Discussions}

\begin{figure} [htbp]
\centering \includegraphics[width=0.94\textwidth]{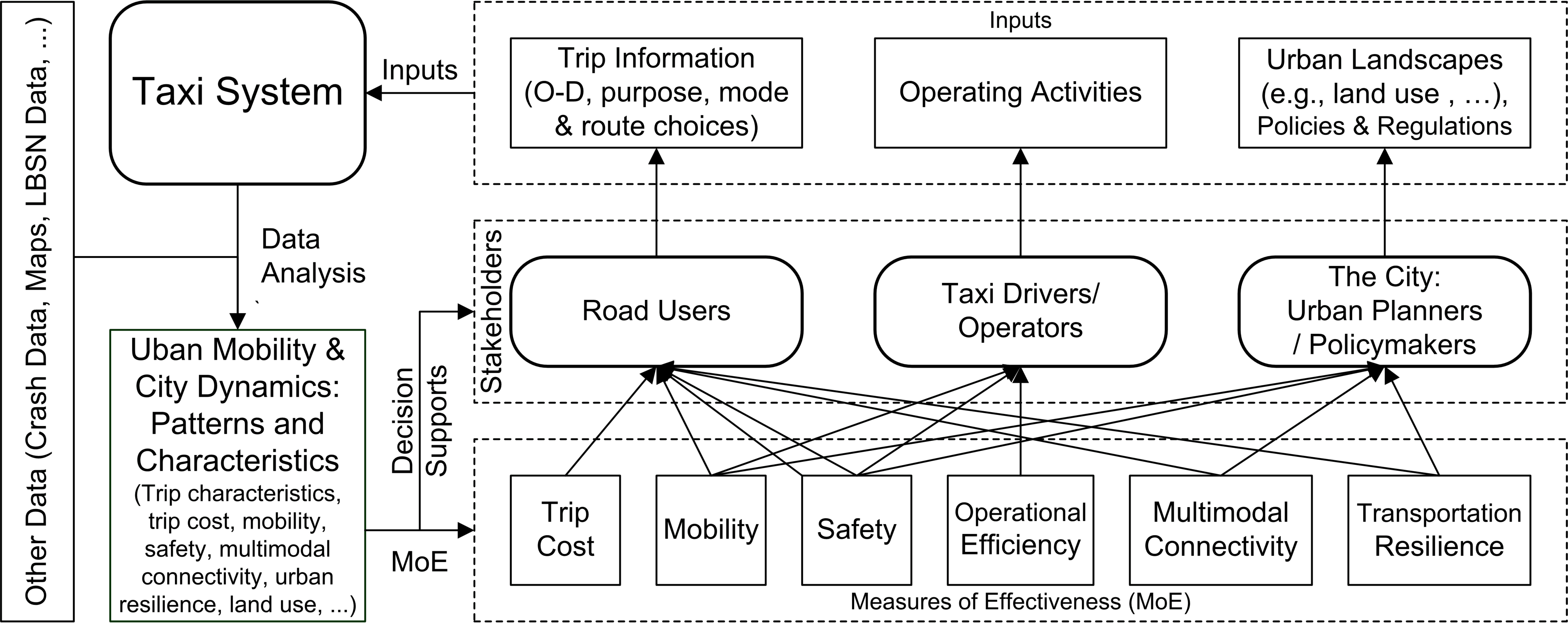} \caption{Data-Driven Decision Supports from Relations Between the Taxi System and Stakeholders.}
\label{fig:BSSStakeholders}
\end{figure}

Here we discuss our work of analyzing taxi O-D trips from the viewpoint of taking advantages of the uncovered patterns and characteristics in urban mobility and city dynamics for data-driven decision supports (DDDS) \citep{Xie2018Bike}. Fig. \ref{fig:BSSStakeholders} shows the DDDS from the relations between the taxi system and key stakeholders such as road users, taxi drivers and operators, and the city (urban planners and policymakers). The inputs from key stakeholders to the taxi system and urban environment include trip information (such as O-D, purpose, mode and route choice), operating strategies (such as pricing \citep{gan2013optimal,yuan2017modeling}, taxi dispatching \citep{miao2016taxi}, and customer searching \citep{tang2016two}), and urban landscapes (e.g., land use and infrastructure) and transportation-related policies and regulations \citep{harding2016taxi}.

For stakeholders, the patterns and characteristics extracted from data can provide key measures of effectiveness (MoE) of the taxi system on an ongoing basis. In this paper, trip cost and mobility are evaluated in Section \ref{sec:mobcost} to measure the impacts of urban traffic congestion. Trip safety and the relations with taxi travel demand are investigated in Section \ref{sec:tripsafety} based on spatial distributions. Notice that although the metrics of mobility and safety represent aggregated experiences from both road users and taxi drivers, they can be improved by city planners or policymakers through upgrading urban landscapes or by taxi operators through adopting emerging technologies for safety such as the innovations on connected and autonomous vehicles. Multimodal connectivity is assessed in Section \ref{sec:connectivity}, where we analyze the current taxi trip demand in regard to other transportation modes including bikeshare and public transit. For road users, a better connectivity can provide them more flexibility and choices in multimodal trip planning. For city planners, more options of multimodal trips can help them make practical policies and regulations to match local needs and conditions and to reduce vehicle miles traveled (VMT) and traffic congestion. 
The role of taxi system on preserving the city-scale resilience is analyzed in Section \ref{sec:resilience}, where we focus on the serious disruptions that caused one or more other major transportation modes to temporarily stop service. The results indicate that taxi system is an important traffic mode retaining resilience for road users. The results are also useful for city policymakers and urban planners on supporting the decision making and planning in emergency events requiring the city-scale resilience.

Beyond providing MoEs, the patterns and characteristics can be used by stakeholders as additional decision supports. The impacts of urban congestion (as shown in Section \ref{sec:mobcost}, measured on cost and mobility ) can serve as evidence-based supports to the city, which for example can be used by urban planners to achieve intelligent transportation network management, and by policymakers to design a more accurate pricing structure \citep{qian2017time,yuan2017modeling} for improving the taxi system efficiency. The relations between trip patterns and land use (explored in Section \ref{sec:landuse}) can be used by taxi system operators to optimize operating activities (such as dispatching and customer searching) and by the city to enhance urban landscapes.

It should be mentioned that although the taxi trip data that we analyzed can only represents a partial group of urban mobility, the findings can contribute to the decision supports for improving overall multimodal urban transportation, in consideration of the similarities between taxi system and other for-hire vehicle services and the complementary relationships between taxi service and other transportation modes.

\section{Conclusion}

In this paper, we performed a systematic analysis to uncover urban mobility and city dynamics in multimodal transportation environments by using the taxi O-D trip data and some additional data sources in the District of Columbia (DC) area. We first studied basic characteristics of taxi trips. Afterwards, we applied data visualization, data analysis, statistical analysis and data fusion to systematically investigate five important aspects, of which three concern urban mobility (respectively mobility and cost including effect of traffic congestion, trip safety, and multimodal connectivity), and the other two are related to city dynamics (respectively transportation resilience, and the relation between trip patterns and land use). For each aspect, we uncovered patterns and characteristics from taxi trip data then explored the results to discuss qualitative and quantitative impacts of the inputs from the stakeholders on available measures of effectiveness on urban mobility and city dynamics.

On characteristics of taxi trips, we found that the statistical distributions of the major taxi trip attributes (including trip distance, time, and fare) follow a parametric lognormal model, which is consistent and complementary with the recent works of the taxi data analysis on a few other different cities. The finding is useful not only for road users by assisting the mode choice in multimodal trip planning, but also for simulation studies by providing inputs to enable evaluation of potential impacts from some major changes in stochastic processes. On mobility and cost, our statistical analysis showed that a majority of taxi trips have a rather low median speed; taxi fare rate drops with the increase in trip distance and finally converges to a cost limit. In addition, the effects of traffic congestion in the city was shown quantitatively measurable from the distributions of trip speed and trip fare of taxi services by time or distance. On road safety, we discussed major safety concerns by analyzing the spatial distribution of taxi crashes and the relationship between crash probability and trip demands. Crash probability was found to follow a common pattern of the scale-free law in the physical and social sciences, where the high-accident locations are subject to a small portion of streets, while a small number of crashes is corresponding to pervasive streets. This means that transportation safety would be greatly improved if the the small group of high-accident streets could receive more safety precautions and efforts, but decreasing crash number to zero would be challenging. High traffic demand was found being a necessary but not a sufficient condition causing a high probability of crashes. On multimodal connectivity, we analyzed the accessibility in connection for taxi passengers to other transportation modes including bikeshare and Metrorail systems. We also discussed the impacts of connection on the upcoming mode shift in a multimodal transportation environment. On transportation resilience, we analyzed two specific cases in DC, in which one is a disruption and another is an extreme event, to study the role of taxi system on preserving resilience in a multimodal transportation environment. The results showed that taxi service system has the capability of fast recovering its function in urban mobility from traffic disruption or extreme event. In case of major disruptions, taxi system was shown to become a crucial transportation mode at the advantages of holding a better essence in retaining resilience compared to the other transportation modes. On the relation between trip patterns and land use, we showed that land use categories can be extracted through analyzing and identifying fundamental trip patterns. The findings can support a better comprehension of city dynamics, especially on the interactions between human mobility dynamics and urban spatial structures.

Finally, we briefly discussed some critical roles and implications of the uncovered patterns and characteristics on urban mobility and city dynamics from the relation between taxi systems and key stakeholders, where key stakeholders include road users, system operators, and city planners and policymakers. The study demonstrated that large-scale statistical analysis on location-based human mobility data has a great value subject to revealing insights of transportation and urban dynamics in a multimodal urban environment. The results in this paper can support road users by providing evidence-based information of trip cost, mobility, safety, multimodal connectivity and transportation resilience, can assist taxi drivers and operators to deliver transportation services in a higher quality of mobility, safety and operational efficiency, and can also help city planners and policy makers to transform multimodal urban transportation in a more effective and efficient way.

\bibliographystyle{chicago}
\bibliography{bib/own_pub,bib/own_ext,bib/traffic,bib/traffic_tlc,bib/lbsn,bib/traffic_emerge,bib/traffic_taxi,bib/traffic_bike,bib/traffic_metro,bib/trafficsafe,bib/stats,bib/sustain,bib/onlineapi,bib/urbansys}

\end{document}